\documentclass[seceq]{ptptex}

\usepackage{graphicx}
\usepackage{wrapft}

\renewcommand{\a}{\alpha}
\renewcommand{\b}{\beta}
\renewcommand{\c}{\gamma}

\newcommand{\e}{\epsilon}

\newcommand{\n}{\nu}
\renewcommand{\t}{\tau}
\newcommand{\z}{\omega}
\newcommand{\G}{\Gamma}

\title{
New Numerical Methods to Evaluate Homogeneous Solutions of 
the Teukolsky Equation II\\
--- Solutions of the Continued Fraction Equation ---
}

\author{
Ryuichi \textsc{Fujita} and Hideyuki \textsc{Tagoshi}%
}

\inst{
Department of Earth and Space Science, Graduate School of Science,\\
Osaka University, Toyonaka 560-0043, Japan\\
}

\abst{
We investigate the solution of the continued fraction equation by which we 
determine ``the renormalized angular momentum parameter'', $\nu$, 
in the formalism developed by Leaver and Mano, Suzuki, and Takasugi.
In this formalism, we describe the 
homogeneous solutions of the radial Teukolsky equation, which is the
basic equation of the black hole perturbation formalism. 
We find that, contrary to the assumption made in previous works,
the solution, $\nu$, becomes complex valued as $\omega$ (the angular frequency)
becomes large for each $l$ and $m$ (the degree and order of the spin-weighted 
spheroidal harmonics). 
We compare the power radiated by gravitational waves from a particle 
in a circular orbit in the equatorial plane around a Kerr black hole 
in two ways, one using the Mano-Suzuki-Takasugi formalism with complex $\nu$ and 
the other using a direct numerical integration method. 
We find that the two methods produce consistent results. 
These facts prove the validity of using complex solutions to determine the
homogeneous solutions of the Teukolsky equation. 
}

\begin{document}

\maketitle

\section{Introduction}
Inspirals of stellar-mass compact objects into a supermassive 
black hole at galactic nuclei are expected to be one of the most important 
sources of gravitational waves for space-based detectors, such 
as the Laser Interferometer Space Antenna (LISA). \cite{LISA,Gair}
Through the observation of gravitational waves from such systems, we may be able to obtain
information regarding the central black hole's spacetime geometry encoded in 
multipole moments. 
We may also obtain astrophysical information, such as
the mass distribution of compact objects in galactic nuclei.
Theoretical waveforms of gravitational waves from such systems are 
required in order to extract such information through data analysis.

To predict the waveforms of extreme mass ratio inspirals, we adopt the 
black hole perturbation approach. In this approach, 
a compact star is treated as a point particle, and the mass of the compact star, $\mu$,
is assumed to be very small compared to the mass of 
the black hole, $M$, i.e. $\mu/M\ll 1$. 
In this context, the Teukolsky equation \cite{Teukolsky}
describes the evolution of a perturbation of the Kerr black hole 
spacetime. The standard approach to solve the Teukolsky equation is 
based on the Green function method. The Green function is constructed 
from two kinds of homogeneous solutions. The solution of the 
Teukolsky equation is obtained by the integration of the Green function
multiplied by the source term. 

In a previous paper \cite{Fujitago} (paper I), 
we discussed a numerical method to evaluate the homogeneous solutions 
using a formalism developed by Mano et al.\cite{MST} (MST).
In this formalism, homogeneous solutions are expressed 
in series of hypergeometric functions around the horizon and 
in series of Coulomb wave functions around infinity.
This formalism is an extension of the formalism of Leaver,\cite{Leaver}
in which only the series of Coulomb wave functions are given. 
One of the most important problems in the application of the Leaver and 
the MST formalisms to numerical computation is 
to determine the so-called ``renormalized angular momentum'', $\nu$, by solving
an equation of the form $g(\nu)=0$, where $g(\nu)$ is expressed in terms of continued fractions.
We found that, for each $s, \ell, m$ and $q$ (where $s$ is the spin index of the Teukolsky 
equation, $\ell$ and $m$ are the indices of the
spin weighted spheroidal harmonics, and $q$ is the Kerr parameter divided by 
the mass of the black hole, $q=a/M$), there is a maximum value of $\omega$ for which
we can obtain $\nu$, if we assume that the $\nu$ is real. 

In this paper, we investigate in more detail the properties of the solution of 
the continued fraction equation $g(\nu)=0$, defined in the Leaver and the MST 
formalisms. 
We search for $\n$ in the complex region if we cannot find $\n$ 
in the real region.
We find that the solution is actually complex valued when 
no real solutions are found. 
Further, we find that the real part of the complex solutions, when they exist, 
always take integer or half-integer values. 
We confirm that these complex solutions of the continued fraction equation
can be used to determine the homogeneous solutions of the Teukolsky equation.
We compare the power radiated by gravitational waves from a particle 
in a circular orbit in the equatorial plane around a Kerr black hole 
computed by the MST formalism with complex $\nu$ 
and by a direct numerical integration method. 
We find that the two methods give consistent results. 
These facts show the validity of using the complex solutions to determine the
homogeneous solutions of the Teukolsky equation. 
By using complex $\n$, we can evaluate gravitational waves 
even in the case that $M\omega$ becomes large. This range of $M\omega$ 
is important for the LISA data analysis which deals with the case 
that a particle moves in the strong gravitational field 
or more generic orbits. 

This paper is organized as follows. 
In $\S$ \ref{sec:MST}, we briefly review the MST formalism and introduce
the continued fraction equation, $g(\nu)=0$.
In $\S$ \ref{sec:NM}, we discuss the properties of the solutions of the 
continued fraction equation. 
In $\S$ \ref{sec:luminosity}, 
we compare the power radiated by gravitational waves from a particle 
around a Kerr black hole 
computed by the MST formalism and by a direct numerical integration method.
Section \ref{sec:summary} is devoted to a summary and discussion. 
Throughout this paper, we use units in which $G=c=1$. 

\section{Analytic solutions of the homogeneous Teukolsky equation, and 
the continued fraction equation}
\label{sec:MST}

In this paper, we consider only the case $\omega\geq 0$. Solutions in the case $\omega<0$
can be obtained from those in the case $\omega>0$ by the symmetry of the Teukolsky equation. 

The homogeneous Teukolsky equation is given by 
\begin{eqnarray}
\label{eq:Teukolsky}
&&\Delta^2{d\over dr}\left({1\over \Delta}{dR_{\ell m\omega}\over dr}
\right)
-V(r) R_{\ell m\omega}=0,
\end{eqnarray}
where the potential term $V(r)$ is given by
\begin{eqnarray}
V(r) = -{K^2 + 4i(r-M)K\over\Delta} + 8i\omega r + \lambda.
\end{eqnarray}
Here $\Delta=r^2-2Mr+a^2=(r-r_+)(r-r_-)$, with $r_\pm=M\pm\sqrt{M^2-a^2}$ and 
$K=(r^2+a^2)\omega-ma$,  and $\lambda$ is the eigenvalue of the angular Teukolsky equation.

We consider the incoming solution of the homogeneous Teukolsky equation defined by 
the asymptotic behavior
\begin{eqnarray}
\label{eq:Rin-asymp}
R_{lm\omega}^{{\rm in}}&\rightarrow&\left\{
\begin{array}{cc}
B_{lm\omega}^{{\rm trans}}\Delta^2 e^{-ikr*}&{\rm for}\ r\rightarrow r_{+},\\
r^3 B_{lm\omega}^{{\rm ref}}e^{i\omega r*}+r^{-1}B_{lm\omega}^{{\rm inc}}e^{-i\z r*}&{\rm for}\ r\rightarrow +\infty,
\end{array}
\right.
\end{eqnarray}
where $k=\omega-ma/2Mr_{+}$ and $r^{*}$ is the tortoise coordinate defined by
\begin{eqnarray}
r^*=r+\frac{2Mr_{+}}{r_{+}-r_{-}}\ln\frac{r-r_{+}}{2M}-\frac{2Mr_{-}}{r_{+}-r_{-}}\ln\frac{r-r_{-}}{2M}.
\end{eqnarray}
Here $r_{\pm}=M\pm\sqrt{M^2-a^2}$. 

In the MST method, the homogeneous solutions of the Teukolsky equation 
are expressed in terms of two kinds of series of special 
functions \cite{MST,MSTR}. One consists of series of hypergeometric 
functions, and the other consists of series of Coulomb wave functions.
The former is convergent at the horizon, while the latter at infinity. 

For the homogeneous solution, $R_{lm\omega}^{{\rm in}}$, we define
$p_{{\rm in}}$ by 
\begin{equation}
R^{{\rm in}}=e^{i\e\kappa x}(-x)^{-s-i(\e+\t)/2}
(1-x)^{i(\e-\t)/2}p_{{\rm in}}(x).
\label{eq:Rin}
\end{equation}
The function $p_{\rm in}$ is expressed in the form of a series of hypergeometric
functions as
\begin{eqnarray}
\label{eq:series of Rin}
p_{{\rm in}}(x)=\displaystyle\sum_{n=-\infty}^{\infty}a_{n}
F(n+\nu+1-i\tau,-n-\n-i\tau;1-s-i\e-i\tau;x),
\end{eqnarray}
where 
$x=\z (r_+ -r)/\e \kappa$, $\e=2M\z, \kappa={\sqrt{1-q^2}}, q=a/M$ and 
$\t={{(\e-mq)}/{\kappa}}$, and $F(\a,\b;\c;x)$ is the hypergeometric function.

Next, we present a solution in the form of a series of Coulomb wave functions. 
Let us denote a homogeneous solution by $R_{{\rm C}}$. 
We define the function $f_{\n}(z)$ through the relation
\begin{equation}
R_{{\rm C}}={z}^{-1-s}\left(1-{\e \kappa \over{{z}}}\right)^{-s-i(\e+\t)/2}
f_{\n}(z). 
\end{equation}
We express the function $f_{\nu}(z)$ in the form of a series of Coulomb wave functions as
\begin{eqnarray}
\label{eq:series of Rc}
f_{\n}(z)= 
\displaystyle\sum_{n=-\infty}^{\infty}
(-i)^n\frac{(\n+1+s-i\e)_n}{(\n+1-s+i\e)_n}a_n F_{n+\n}(-is-\e,z),
\end{eqnarray}
where $z=\z (r- r_- )$ and $(a)_{n}=\Gamma(a+n)/\Gamma(a)$,
and $F_{N}(\eta,z)$ is a Coulomb wave function defined by 
\begin{eqnarray}
F_{N}(\eta,z)=e^{-iz}2^{N}z^{N+1}\frac{\G(N+1-i\eta)}{\G(2N+2)}\Phi(N+1-i\eta,
2N+2;2iz).
\label{eq:defcoulomb}
\end{eqnarray}
Here $\Phi(\a,\b;z)$ is the confluent hypergeometric function, 
which is regular at $z=0$ (see $\S$ 13 of Ref.~\citen{handbook}). 
The function $R_{{\rm C}}$ is related to $R^{{\rm in}}$ by
\begin{equation}
R^{{\rm in}}=K_\nu R^\nu_{\rm C}+K_{-\nu-1} R^{-\nu-1}_{\rm C},
\end{equation}
where $K_\nu$ is given by Eq. (2$\cdot$20) of paper I. 

The expansion coefficients 
of the series of hypergeometric functions and the series of Coulomb
wave functions, $\{a_n\}$, satisfy the same three-term recurrence relation,
given by 
\begin{eqnarray}
\label{eq:3term}
\alpha_n^\nu a_{n+1}+\beta_n^{\nu} a_{n}+\gamma_n^\nu a_{n-1}=0,
\end{eqnarray}
where
\begin{eqnarray}
\a_n^\n&=&i\e \kappa (n+\n)(2n+2\n-1)(n+\n+1+s+i\e)\nonumber\\
&&\quad\quad \times(n+\n+1+s-i\e)(n+\n+1+i\t),
\label{eq:def_alpha}
\\
\b_n^\n&=&\left(-\lambda-s(s+1)+(n+\n)(n+\n+1)+\e^2+\e(\e-mq)\right)\nonumber\\
&&\quad\quad\times (n+\n)(n+\n+1)(2n+2\n+3)(2n+2\n-1)
\nonumber\\
&&\quad
+\e (\e-mq)(s^2+\e^2)(2n+2\n+3)(2n+2\n-1),
\label{eq:def_beta}
\\
\c_n^\n&=&-i\e \kappa (n+\n+1)(2n+2\n+3)(n+\n-s+i\e)\nonumber\\
&&\quad\quad \times(n+\n-s-i\e)(n+\n-i\t).
\label{eq:def_gamma}
\end{eqnarray}
The definitions of $\a_n^\n, \b_n^\n$ and $\c_n^\n$ are different 
from those given in paper I by the factor $(n+\n)(n+\n+1)(2n+2\n+3)(2n+2\n-1)$.
The series of Coulomb wave functions and 
the three term recurrence relation Eq. (\ref{eq:3term})
are equivalent to those obtained by Leaver\cite{Leaver}. 

We note that the parameter $\nu$ introduced in the above formulas
does not exist in the Teukolsky equation. 
This parameter is introduced so that both series 
converge and actually represent a solution of the Teukolsky equation. 
Here, we review the method to determine $\nu$ and obtain the convergence 
of both series expansions, based on Ref.~\citen{ST}. 

There are two independent solutions of Eq. (\ref{eq:3term}) whose 
asymptotic behavior for $n\rightarrow\infty$ are given by 
\begin{eqnarray}
\lim_{n\rightarrow\infty} n\frac{a_n^{(1)}}{a_{n-1}^{(2)}}=\frac{i\epsilon\kappa}{2},
\quad
\lim_{n\rightarrow\infty} \frac{a_n^{(2)}}{na_{n-1}^{(2)}}=\frac{2i}{\epsilon\kappa}.
\end{eqnarray}
There are also two independent solutions whose asymptotic behavior 
for $n\rightarrow -\infty$ are given by 
\begin{eqnarray}
\lim_{n\rightarrow -\infty} n\frac{a_n^{(3)}}{a_{n-1}^{(3)}}=-\frac{i\epsilon\kappa}{2},
\quad
\lim_{n\rightarrow -\infty} \frac{a_n^{(4)}}{na_{n-1}^{(4)}}=-\frac{2i}{\epsilon\kappa}. 
\end{eqnarray}
It is shown that in order to have convergence of Eqs. (\ref{eq:series of Rin})
and (\ref{eq:series of Rc}), the coefficients, $a_n^{(1)}$, must be used for $n\rightarrow\infty$,
and the coefficients $a_n^{(3)}$ must be used for $n\rightarrow -\infty$. 
These two solutions are called the ``minimal solution'' as $n\rightarrow \pm\infty$,
respectively. 

In fact, it can be proved that if we use these solutions 
as the expansion coefficients $\{a_n\}$, the series of hypergeometric
functions Eq. (\ref{eq:series of Rin}) converges for $x$ in the range 
$-\infty< x\leq 0$, and the series Eq. (\ref{eq:series of Rc}) converges
for $z>\epsilon\kappa$, or equivalently, $r>r_+$. 
However, these two solutions 
do not coincide in general. 
In order to obtain a consistent solution, 
the parameter $\nu$ is used. 

We next introduce the following quantities:
\begin{equation}
R_n\equiv {a_{n}^{(1)}\over a_{n-1}^{(1)}}\,,\quad
L_n\equiv {a_{n}^{(3)}\over a_{n+1}^{(3)}}\,.
\end{equation}
We can express $R_{n}$ and $L_{n}$ in terms of continued fractions as 
\begin{eqnarray}
R_n&=&-{\gamma_n^\nu\over {\beta_n^\nu+\alpha_n^\nu R_{n+1}}}
\nonumber\\
&=&-{\gamma_{n}^\nu\over \beta_{n}^\nu-}\,
{\alpha_{n}^\n\gamma_{n+1}^\nu\over \beta_{n+1}^\nu-}\,
{\alpha_{n+1}^\n\gamma_{n+2}^\nu\over \beta_{n+2}^\nu-}\cdots,
\label{eq:Rncont}\\
L_n&=&-{\alpha_n^\nu\over {\beta_n^\nu+\gamma_n^\nu L_{n-1}}}
\nonumber\\
&=&-{\alpha_{n}^\nu\over \beta_{n}^\nu-}\,
{\alpha_{n-1}^\n\gamma_{n}^\nu\over \beta_{n-1}^\nu-}\,
{\alpha_{n-2}^\n\gamma_{n-1}^\nu\over \beta_{n-2}^\nu-}\cdots.
\label{eq:Lncont}
\end{eqnarray}
The convergence of the above continued fractions is guaranteed, because $\{a_n^{(1)}\}$ 
and $\{a_n^{(3)}\}$ constitute minimal solutions\cite{Gautschi}. 

In order to have the coincidence of the two solutions $\{a_n^{(1)}\}$ and $\{a_n^{(3)}\}$,
we impose the relation
\begin{eqnarray}
\label{eq:consistency}
R_{n}L_{n-1}=1.
\end{eqnarray}
If we choose $\nu$ such that this implicit equation is satisfied
for a certain $n$, we can obtain a minimal solution
$\{a_n\}$ that is valid over the entire range $-\infty<n<\infty$. 

In place of Eq. (\ref{eq:consistency}), 
we define an equivalent equation.
First we divide Eq. (\ref{eq:3term}) by $a_{n}$ and replace $a_n/a_{n-1}$ by $R_n$
and $a_n/a_{n+1}$ by $L_n$.  We then find 
\begin{eqnarray}
\label{eq:determine_nu}
g_n(\nu)\equiv \beta_n^{\nu}+\alpha_n^\nu R_{n+1}+\gamma_n^\nu L_{n-1}=0,
\end{eqnarray}
where $R_{n+1}$ and $L_{n-1}$ are given by the continued
fractions Eq. (\ref{eq:Rncont}) and Eq. (\ref{eq:Lncont}), respectively. 
Equation (\ref{eq:determine_nu}) is a transcendental equation expressed in terms of  
continued fractions. We call Eq. (\ref{eq:determine_nu}) the ``continued fraction equation''
in this paper. 

\section{Properties of the solutions of the continued fraction equation}
\label{sec:NM}

We first note that $g_n(\nu)$ in Eq. (\ref{eq:determine_nu}) is invariant under 
the simultaneous transformation $\nu\leftrightarrow -\nu-1$ and $n\leftrightarrow -n$. 
Thus, $g_0(\nu)$ is symmetric under $\nu\leftrightarrow -\nu-1$.
In particular, if we plot $g_0(\nu)$ as a function of $\nu$, 
$g_0(\nu)$ is symmetric about the line $\nu=-1/2$ in the complex plane. 

Next, we note that 
the coefficients $\alpha_n^\nu$, $\beta_n^\nu$
and $\gamma_n^\nu$ in Eqs. (\ref{eq:def_alpha}) -- (\ref{eq:def_gamma}) contain
$\nu$ only in the form $n+\nu$.
This implies that $g_n(\nu+1)=g_{n+1}(\nu)$. 
When $\nu$ is a solution of Eq. (\ref{eq:determine_nu}), we have $g_n(\nu)=0$. 
This can be rewritten as 
\begin{eqnarray*}
g_{n+1}(\nu)=\beta_{n+1}+\alpha_{n+1}R_{n+2}+\gamma_{n+1} L_n=0. 
\end{eqnarray*}
We then have $0=g_{n+1}(\nu)=g_n(\nu+1)$. 
Thus, $\nu+1$ is also a solution. 
Namely, if $g_n(\nu)=0$, $g_n(\nu+k)=0$ for an arbitrary integer $k$.
These considerations show that 
it is sufficient to investigate $g_0(\nu)=0$ to find the solution $\nu$. 

\begin{figure}[t]
\parbox{\halftext}{
\centerline{\includegraphics[width=6.5cm,height=6cm]{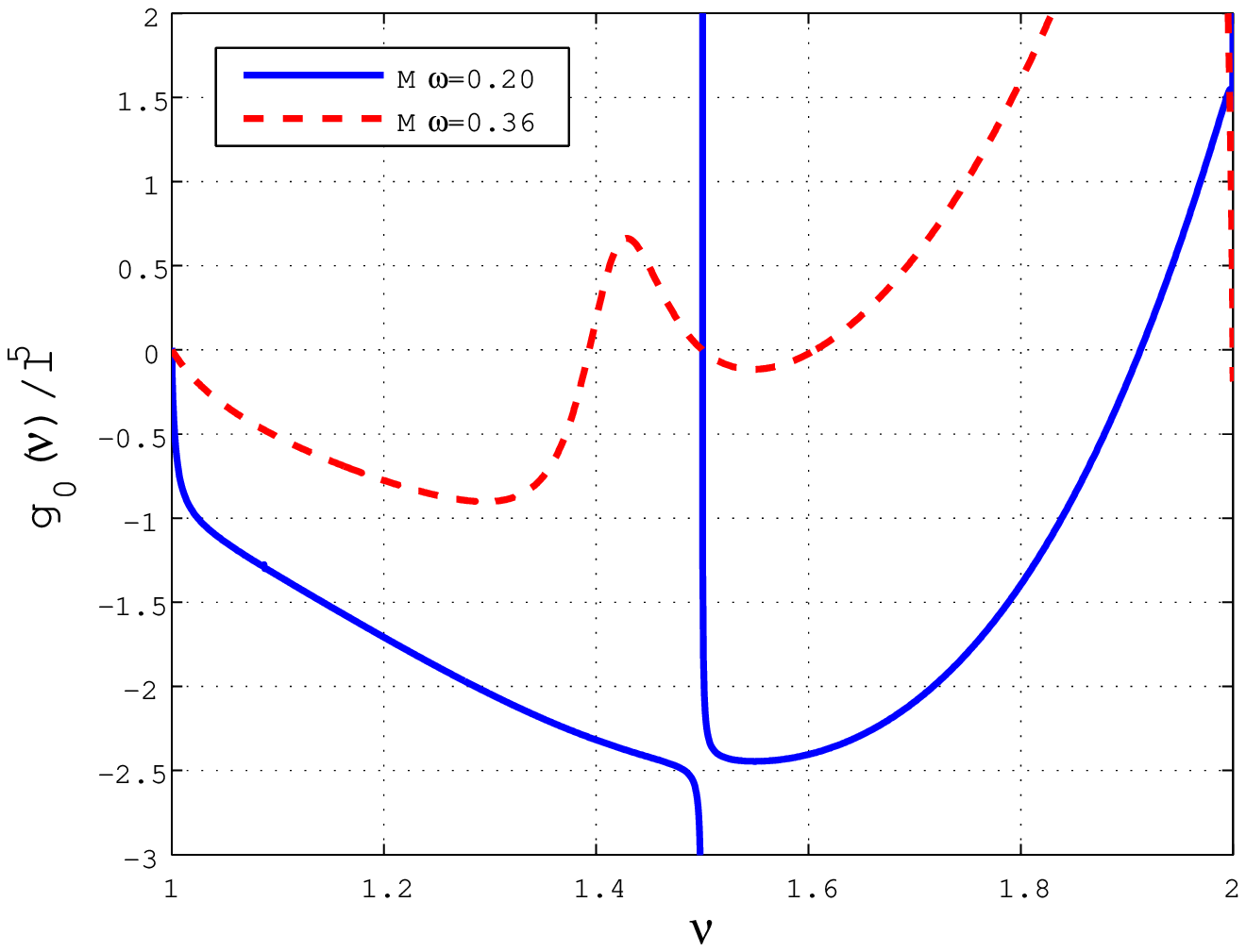}}
\caption{The plot of $g_0(x)/l^5$ for $M\omega=0.20$ and 0.36. Here $s=-2$,
$l=m=2$ and $q=0$. 
There exist real solutions in these cases. }\label{fig:funcnu1}
}
\hfill
\parbox{\halftext}{
\centerline{\includegraphics[width=6.5cm,height=6cm]{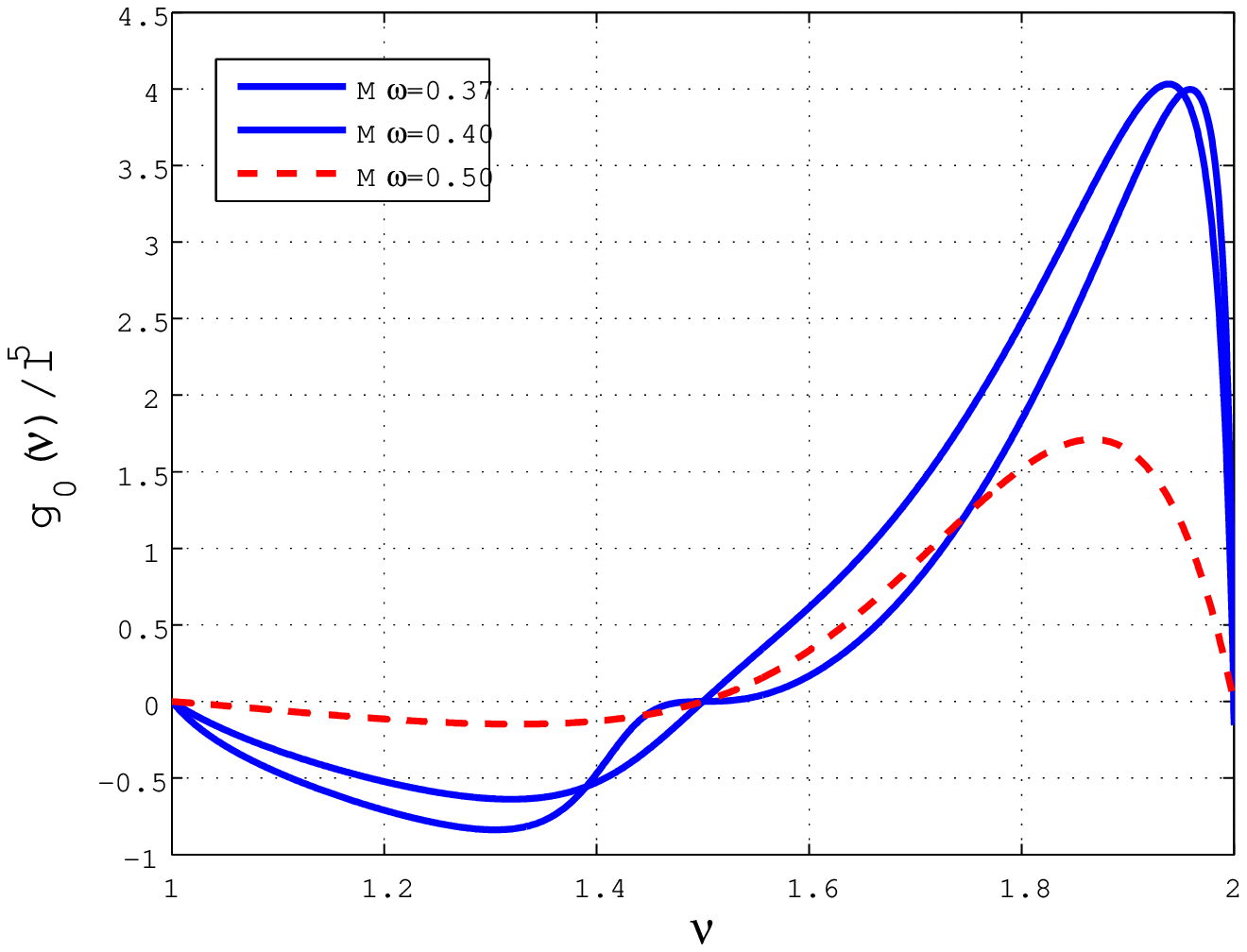}}
\caption{The plot of $g_0(x)/l^5$ for $M\omega=0.37$, 0.40 and 0.50. Here $s=-2$,
$l=m=2$ and $q=0$. 
No real solutions are found in these cases. }\label{fig:funcnu2}
}
\end{figure}

When $\epsilon=2M\omega$ is small, there is an analytic expression 
of a solution $\nu$ in the form of a series in $\epsilon$, given by 
\begin{eqnarray}
\label{eq:nusol}
\nu&=& l+{1\over{2l+1}}\left[-2-{s^2\over{l(l+1)}}
+{[(l+1)^2-s^2]^2
\over{(2l+1)(2l+2)(2l+3)}}
\right.
\nonumber\\
&&
\left.
-{(l^2-s^2)^2\over{(2l-1)2l(2l+1)}}\right]
\epsilon^2+O(\epsilon^3). 
\end{eqnarray}
Thus, when $\epsilon\rightarrow 0$, $\nu$ takes an integer value
$\nu=l+k$ (where $k$ is an arbitrary integer). 
As an example, we consider the case $l=m=2$, $q=0$ and $k=0$. 
Then, as $\omega$ increases from 0, the solution $\nu$ decreases from $l$ 
and approaches $l-1/2$ 
near $M\omega=0.36$. If $M\omega$ is larger than $0.36$, 
the real solution $\nu$ disappears. The plot of $g_0(x)/l^5$ 
is shown in Fig. \ref{fig:funcnu1} in the case $M\omega\leq 0.36$ 
and in Fig. \ref{fig:funcnu2} in the case $M\omega>0.36$. 
The factor $l^{-5}$ is introduced so that the function $g_0(x)/l^5$ remains order unity. 
At first glance, 
in Fig. \ref{fig:funcnu2}, there seems to be a solution at $x=l-1/2=3/2$, $x=1$ or $x=2$. 
However, at these points, $g_0(x)$ always becomes 0, irrespective of $\omega$. 
These are solutions of $g_0(x)=0$ mathematically, but they are not solutions
that can be used to connect two minimal solutions in the limits $n=\pm\infty$, discussed in the previous section. 
In Table I of paper I, we listed the maximum values of $M\omega$ 
for which real $\nu$ is found. We also list them in Table \ref{tab:maximum_omega} 
for convenience. We express the maximum value of $M\omega$ 
as $(M\omega)_{\rm max}$. 

{\tiny
\begin{table}[htbp]
\caption{Maximum values of $M\omega$ for which real solutions $\nu$ are found.}
\begin{center}
\begin{tabular}{cc|ccc||cc|ccc}
\hline \hline 
$\ell$&$m$&$q=0.9$&$q=0$&$q=-0.9$&
$\ell$&$m$&$q=0.9$&$q=0$&$q=-0.9$
\\ \hline 
2 & 2  &0.39 &0.36 & 0.32 & 8 & 8  &1.08 &1.00 & 0.92 \\
2 & 1  &0.44 &0.36 & 0.34 & 8 & 7  &1.07 &1.00 & 0.93 \\
2 & 0  &0.38 &0.36 & 0.38 & 8 & 6  &1.06 &1.00 & 0.94 \\
3 & 3  &0.58 &0.53 & 0.45 & 8 & 5  &1.05 &1.00 & 0.95 \\
3 & 2  &0.61 &0.53 & 0.48 & 8 & 4  &1.04 &1.00 & 0.96 \\
3 & 1  &0.59 &0.53 & 0.51 & 8 & 3  &1.03 &1.00 & 0.97 \\
3 & 0  &0.55 &0.53 & 0.55 & 8 & 2  &1.02 &1.00 & 0.98 \\
4 & 4  &0.73 &0.66 & 0.57 & 8 & 1  &1.01 &1.00 & 0.99 \\
4 & 3  &0.73 &0.66 & 0.60 & 8 & 0  &1.00 &1.00 & 1.00 \\
4 & 2  &0.72 &0.66 & 0.63 & 9 & 9  &1.14 &1.06 & 0.99 \\
4 & 1  &0.70 &0.66 & 0.65 & 9 & 8  &1.13 &1.06 & 1.00 \\
4 & 0  &0.68 &0.66 & 0.68 & 9 & 7  &1.12 &1.06 & 1.00 \\
5 & 5  &0.84 &0.77 & 0.68 & 9 & 6  &1.12 &1.06 & 1.01 \\
5 & 4  &0.83 &0.77 & 0.70 & 9 & 5  &1.11 &1.06 & 1.02 \\
5 & 3  &0.82 &0.77 & 0.72 & 9 & 4  &1.10 &1.06 & 1.03 \\
5 & 2  &0.80 &0.77 & 0.74 & 9 & 3  &1.09 &1.06 & 1.04 \\
5 & 1  &0.79 &0.77 & 0.76 & 9 & 2  &1.08 &1.06 & 1.05 \\
5 & 0  &0.77 &0.77 & 0.77 & 9 & 1  &1.07 &1.06 & 1.06 \\
6 & 6  &0.93 &0.85 & 0.77 & 9 & 0  &1.06 &1.06 & 1.06 \\
6 & 5  &0.92 &0.85 & 0.79 & 10& 10 &1.20 &1.12 & 1.05 \\
6 & 4  &0.91 &0.85 & 0.80 & 10& 9  &1.19 &1.12 & 1.05 \\
6 & 3  &0.89 &0.85 & 0.82 & 10& 8  &1.19 &1.12 & 1.06 \\
6 & 2  &0.88 &0.85 & 0.83 & 10& 7  &1.18 &1.12 & 1.07 \\
6 & 1  &0.87 &0.85 & 0.84 & 10& 6  &1.17 &1.12 & 1.08 \\
6 & 0  &0.86 &0.85 & 0.86 & 10& 5  &1.16 &1.12 & 1.09 \\
7 & 7  &1.01 &0.93 & 0.85 & 10& 4  &1.16 &1.12 & 1.09 \\
7 & 6  &1.00 &0.93 & 0.86 & 10& 3  &1.15 &1.12 & 1.10 \\
7 & 5  &0.99 &0.93 & 0.88 & 10& 2  &1.14 &1.12 & 1.11 \\
7 & 4  &0.98 &0.93 & 0.89 & 10& 1  &1.13 &1.12 & 1.12 \\
7 & 3  &0.96 &0.93 & 0.90 & 10& 0  &1.12 &1.12 & 1.12 \\
7 & 2  &0.95 &0.93 & 0.91 &   &    &     &     &      \\
7 & 1  &0.94 &0.93 & 0.92 &   &    &     &     &      \\
7 & 0  &0.93 &0.93 & 0.93 &   &    &     &     &      \\
\hline \hline
\end{tabular}
\end{center}
\label{tab:maximum_omega}
\end{table}
}

\begin{figure}[t]
\parbox{\halftext}{
\centerline{\includegraphics[width=6.5cm,height=6cm]{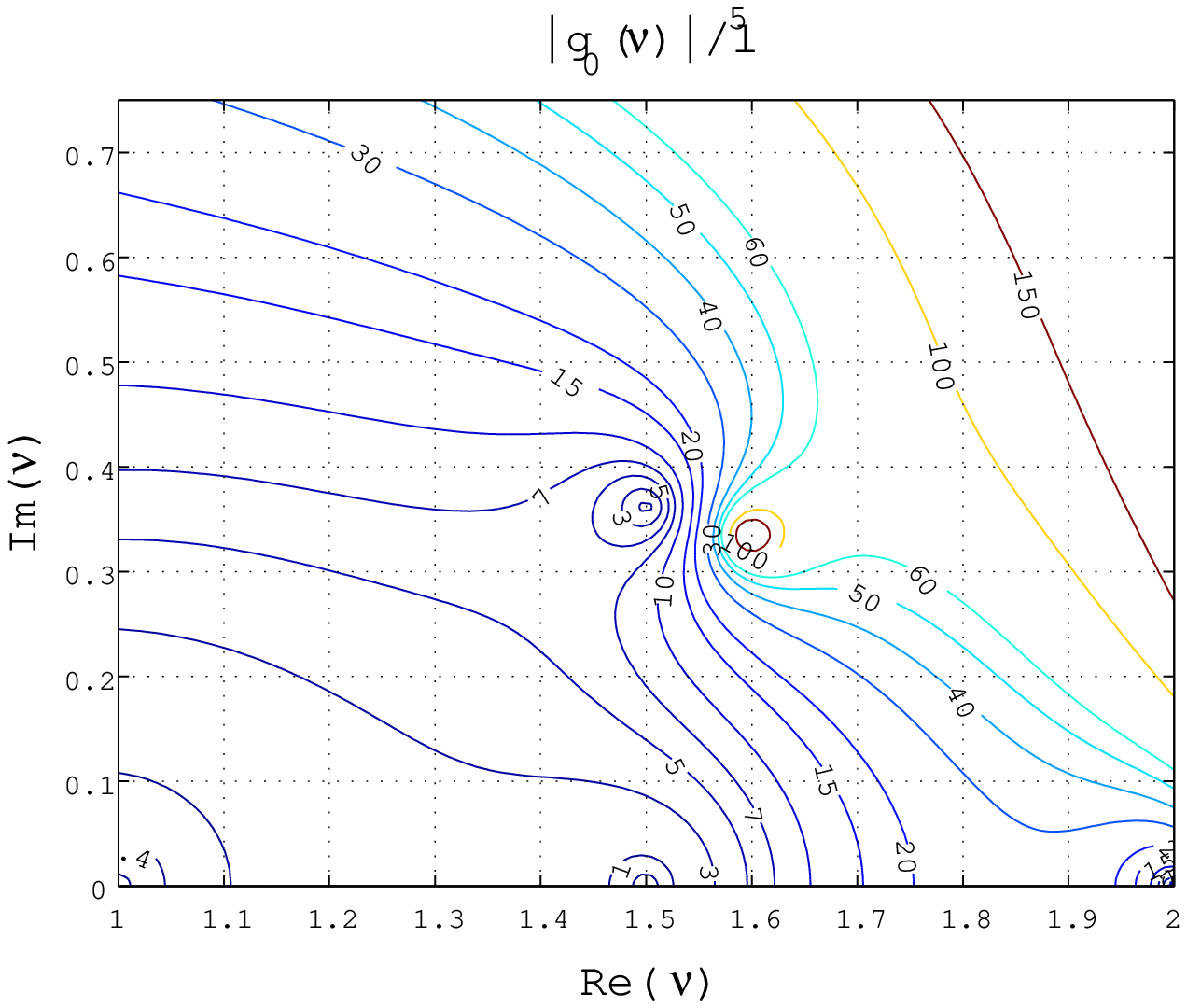}}
\caption{The contour plot of $|g_0(z)|/l^5$ in the case $M\omega=0.50$, $s=-2$, $l=m=2$ and $q=0$.
There is a complex solution at $\nu=1.5+0.36188061539416 i$. }
\label{fig:cfuncnu1}
}
\hfill
\parbox{\halftext}{
\centerline{\includegraphics[width=6.5cm,height=6cm]{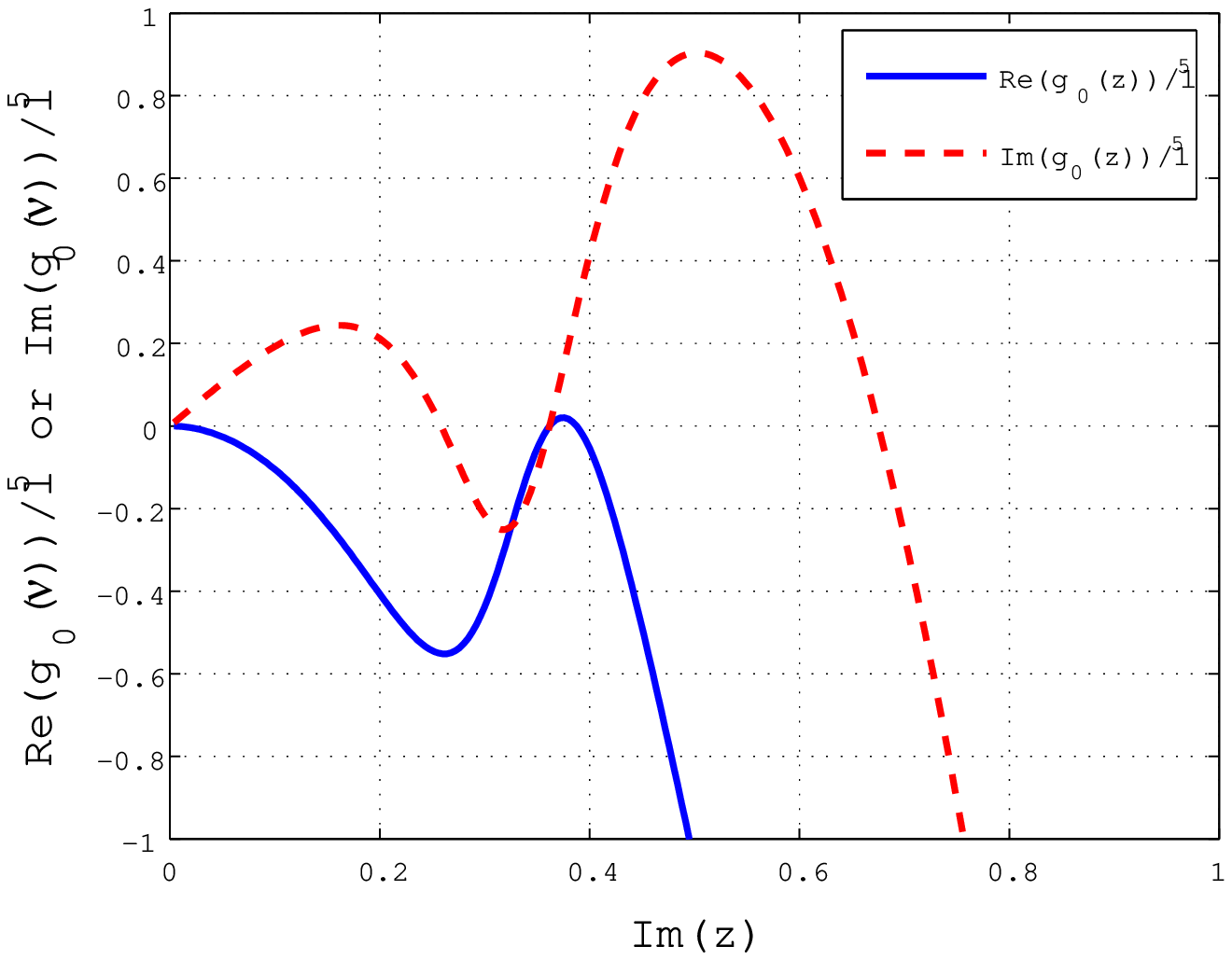}}
\caption{The real and imaginary parts of the function $g_0(z)/l^5$ 
at ${\rm Re}(z)=1.5$ 
in the case $M\omega=0.50$, $s=-2$, $l=m=2$ and $q=0$.
The real and imaginary parts become 0 simultaneously at ${\rm Im}(z)=0.36188061539416$.
Thus, this function satisfies $g_0(z)=0$. }
\label{fig:funcnu-RI}
}
\end{figure}

Next we consider the solution of $g_0(z)=0$ in the complex $z$ plane
in the case that $M\omega$ is larger than $(M\omega)_{\rm max}$ and we cannot find a real solution. 
In Fig. \ref{fig:cfuncnu1}, we plot a contour of $|g_0(z)|$ in the case $M\omega=0.5$. 
We find that there is a minimum near ${\rm Re}(z)=3/2$, ${\rm Im}(z)=0.36$. 
In Fig. \ref{fig:funcnu-RI}, we plot ${\rm Re}(g_0(z))$ and ${\rm Im}(g_0(z))$ as functions of
${\rm Im}(z)$ at ${\rm Re}(z)=3/2$. 
It is evident that there is a solution near $z=3/2+0.36i$. 
The precise value of the solution is $z=3/2+0.36188061539416 i$. 
It is also suggested by Fig. \ref{fig:cfuncnu1} that there are no solutions 
other than this value. We find that when there are no real solutions,
a complex solution always exists. Further, the real part of the complex solution
is always a half-integer or integer. These properties are the same when 
the parameters, $s,l,m$ and $q$ have different values. 

We consider the case in which $M\omega$ is much larger than $(M\omega)_{\rm max}$. 
In Table \ref{tab:nu-l2m2q0}, we list numerical values of the solution $\nu$ 
in the case $s=-2$, $l=m=2$ and $q=0$. 
We find that ${\rm Im}(\nu)$ increases as $M\omega$ increases, 
and it reaches a maximum value near $M\omega=0.53$. 
Then, ${\rm Im}(\nu)$ begins to decrease and approaches 0, 
and the real solution appears again at $M\omega=0.594$. 
This new real solution starts at the half integer value $\nu=3/2$, 
and it approaches the integer value $\nu=2$. 
At $M\omega=0.604$, where the real solution reaches $\nu=2$, 
the real solution disappears, and the imaginary part appears again. 
In Table \ref{tab:nu-l2m2q099}, we list numerical values of the solution $\nu$
in the case $s=-2$, $l=m=2$ and $q=\pm 0.99$. 
When $q=0.99$, the real solution disappears at $M\omega=0.39$, and a complex solution
with ${\rm Re}(\nu)=3/2$ appears. Then ${\rm Im}(\nu)$ continues to increase
until $M\omega=3.0$. 
When $q=-0.99$, the real solution disappears at $M\omega=0.33$, and a complex solution
with ${\rm Re}(\nu)=3/2$ appears. Then ${\rm Im}(\nu)$ continues to increase
until $M\omega=3.0$. 
In Tables \ref{tab:nu-l2m2q0} and \ref{tab:nu-l2m2q099}, we list values for the case $M\omega\leq 3.0$.
Above this value, it becomes difficult to evaluate the continued fraction 
equation numerically, because of the loss of precision.
This happens because as $M\omega$ becomes large, $\beta_n$ approaches 
$-(\alpha_n^\nu R_{n+1}+\gamma_n^\nu L_{n-1})$ irrespective of $\nu$. 

{\tiny
\begin{table}[t]
\caption{Table of $\nu$ for various values of $M\omega$ in the case $s=-2$, $l=m=2$ and $q=0$. }
\begin{center}
\begin{tabular}{c|ll}
\hline \hline 
$M\omega$ & ${\rm Re}(\nu)$ & ${\rm Im}(\nu)$ \\ \hline 
0.1000 & 1.9793154547208 & 0.0000000000000 \\
0.2000 & 1.9129832302687 & 0.0000000000000 \\
0.3000 & 1.7792805424199 & 0.0000000000000 \\
0.3100 & 1.7594378913945 & 0.0000000000000 \\
0.3200 & 1.7375329132632 & 0.0000000000000 \\
0.3300 & 1.7129796083608 & 0.0000000000000 \\
0.3400 & 1.6847877144372 & 0.0000000000000 \\
0.3500 & 1.6510165965479 & 0.0000000000000 \\
0.3600 & 1.6066167352463 & 0.0000000000000 \\
0.3700 & 1.5000000000000 & 0.0117805002266 \\
0.3800 & 1.5000000000000 & 0.1080591170505 \\
0.4000 & 1.5000000000000 & 0.1862468531447 \\
0.4200 & 1.5000000000000 & 0.2393542510924 \\
0.4400 & 1.5000000000000 & 0.2810676246093 \\
0.4600 & 1.5000000000000 & 0.3149109523037 \\
0.4800 & 1.5000000000000 & 0.3419264795996 \\
0.5000 & 1.5000000000000 & 0.3618806153941 \\
0.5200 & 1.5000000000000 & 0.3732918843902 \\
0.5300 & 1.5000000000000 & 0.3748108599871 \\
0.5400 & 1.5000000000000 & 0.3725753632958 \\
0.5600 & 1.5000000000000 & 0.3508153681782 \\
0.5800 & 1.5000000000000 & 0.2767032401575 \\
0.5920 & 1.5000000000000 & 0.1112238120363 \\
0.5930 & 1.5000000000000 & 0.0654237590662 \\
0.5940 & 1.5671748369160 & 0.0000000000000 \\
0.5960 & 1.6572267840117 & 0.0000000000000 \\
0.5980 & 1.7231093132436 & 0.0000000000000 \\
0.6000 & 1.7878302655744 & 0.0000000000000 \\
0.6020 & 1.8649828581648 & 0.0000000000000 \\
0.6030 & 1.9221344236998 & 0.0000000000000 \\
0.6040 & 2.0000000000000 & 0.0739848339575 \\
0.6060 & 2.0000000000000 & 0.1644030708573 \\
0.6080 & 2.0000000000000 & 0.2167770350574 \\
0.7000 & 2.0000000000000 & 0.8003377636925 \\
0.8000 & 2.0000000000000 & 1.1099466644118 \\
0.9000 & 2.0000000000000 & 1.3699138540831 \\
1.0000 & 2.0000000000000 & 1.6085538776570 \\
1.2000 & 2.0000000000000 & 2.0541150948294 \\
1.4000 & 2.0000000000000 & 2.4772658891494 \\
1.6000 & 2.0000000000000 & 2.8879097347262 \\
1.8000 & 2.0000000000000 & 3.2903727081665 \\
2.0000 & 2.0000000000000 & 3.6867890278893 \\
2.2000 & 2.0000000000000 & 4.0782125805902 \\
2.4000 & 2.0000000000000 & 4.4650426712959 \\
2.6000 & 2.0000000000000 & 4.8471678988185 \\
2.8000 & 2.0000000000000 & 5.2239336390821 \\
3.0000 & 2.0000000000000 & 5.5939000509184 \\
\hline
\end{tabular}
\end{center}
\label{tab:nu-l2m2q0}
\end{table}
}

{\tiny
\begin{table}[t]
\caption{Table of $\nu$ for various values of $M\omega$ in the case $q=0.99$ and $-0.99$, $s=-2$, $l=m=2$. }
\begin{center}
\begin{tabular}{c|ll|ll}
\hline \hline 
& \multicolumn{2}{|c|}{$q=0.99$} & \multicolumn{2}{|c}{$q=-0.99$} \\ \hline
$M\omega$ & ${\rm Re}(\nu)$ & ${\rm Im}(\nu)$ & ${\rm Re}(\nu)$ & ${\rm Im}(\nu)$ \\ \hline 
 0.10 & 1.9806255500899 & 0.0000000000000                 & 1.9778681909419 & 0.0000000000000 \\
 0.20 & 1.9239618344032 & 0.0000000000000                 & 1.8984257752783 & 0.0000000000000 \\
 0.30 & 1.8183581833078 & 0.0000000000000                 & 1.6857218860079 & 0.0000000000000 \\
 0.31 & 1.8027568870615 & 0.0000000000000                 & 1.6371352569052 & 0.0000000000000 \\
 0.32 & 1.7854692988454 & 0.0000000000000                 & 1.5499916877959 & 0.0000000000000 \\
 0.33 & 1.7660512947108 & 0.0000000000000                 & 1.5000000000000 & 0.1201460888531 \\
 0.34 & 1.7438484200603 & 0.0000000000000                 & 1.5000000000000 & 0.1788957377084 \\
 0.35 & 1.7178247791233 & 0.0000000000000                 & 1.5000000000000 & 0.2240840951862 \\
 0.36 & 1.6861420552899 & 0.0000000000000                 & 1.5000000000000 & 0.2628158063725 \\
 0.37 & 1.6447942140179 & 0.0000000000000                 & 1.5000000000000 & 0.2976343179397 \\
 0.38 & 1.5793009665553 & 0.0000000000000                 & 1.5000000000000 & 0.3297922470347 \\
 0.39 & 1.5000000000000 & 0.0974489094430                 & 1.5000000000000 & 0.3600132102046 \\
 0.40 & 1.5000000000000 & 0.1628152952473                 & 1.5000000000000 & 0.3887584134857 \\
 0.50 & 1.5000000000000 & 0.5368498162041                 & 1.5000000000000 & 0.6365360949922 \\
 0.60 & 1.5000000000000 & 0.8672067267975                 & 1.5000000000000 & 0.8571228112204 \\
 0.70 & 1.5000000000000 & 1.1838651090370                 & 1.5000000000000 & 1.0704661825964 \\
 0.80 & 1.5000000000000 & 1.4840266489876                 & 1.5000000000000 & 1.2821245153699 \\
 0.90 & 1.5000000000000 & 1.7701386563708                 & 1.5000000000000 & 1.4938849397247 \\
 1.00 & 1.5000000000000 & 2.0454330629622                 & 1.5000000000000 & 1.7061371538867 \\
 1.10 & 1.5000000000000 & 2.3125401729619                 & 1.5000000000000 & 1.9187278288162 \\
 1.20 & 1.5000000000000 & 2.5734029380085                 & 1.5000000000000 & 2.1313317677628 \\
 1.30 & 1.5000000000000 & 2.8294329783970                 & 1.5000000000000 & 2.3436142271920 \\
 1.40 & 1.5000000000000 & 3.0816629547228                 & 1.5000000000000 & 2.5552922279595 \\
 1.50 & 1.5000000000000 & 3.3308588672551                 & 1.5000000000000 & 2.7661482256677 \\
 1.60 & 1.5000000000000 & 3.5775977157128                 & 1.5000000000000 & 2.9760238443012 \\
 1.70 & 1.5000000000000 & 3.8223206679586                 & 1.5000000000000 & 3.1848072932141 \\
 1.80 & 1.5000000000000 & 4.0653697494531                 & 1.5000000000000 & 3.3924204618486 \\
 1.90 & 1.5000000000000 & 4.3070135210021                 & 1.5000000000000 & 3.5988078524059 \\
 2.00 & 1.5000000000000 & 4.5474653357747                 & 1.5000000000000 & 3.8039277638612 \\
 2.10 & 1.5000000000000 & 4.7868965237373                 & 1.5000000000000 & 4.0077454471779 \\
 2.20 & 1.5000000000000 & 5.0254460525305                 & 1.5000000000000 & 4.2102277404164 \\
 2.30 & 1.5000000000000 & 5.2632277012574                 & 1.5000000000000 & 4.4113386736802 \\
 2.40 & 1.5000000000000 & 5.5003354519402                 & 1.5000000000000 & 4.6110355696752 \\
 2.50 & 1.5000000000000 & 5.7368475857391                 & 1.5000000000000 & 4.8092651925867 \\
 2.60 & 1.5000000000000 & 5.9728298259083                 & 1.5000000000000 & 5.0059594834575 \\
 2.70 & 1.5000000000000 & 6.2083377711225                 & 1.5000000000000 & 5.2010303381768 \\
 2.80 & 1.5000000000000 & 6.4434187950814                 & 1.5000000000000 & 5.3943626971054 \\
 2.90 & 1.5000000000000 & 6.6781135409542                 & 1.5000000000000 & 5.5858048534288 \\
 3.00 & 1.5000000000000 & 6.9124571056503                 & 1.5000000000000 & 5.7751542098469 \\
\hline
\end{tabular}
\end{center}
\label{tab:nu-l2m2q099}
\end{table}
}

\begin{figure}[t]
\parbox{\halftext}{
\centerline{\includegraphics[width=6.5cm,height=6cm]{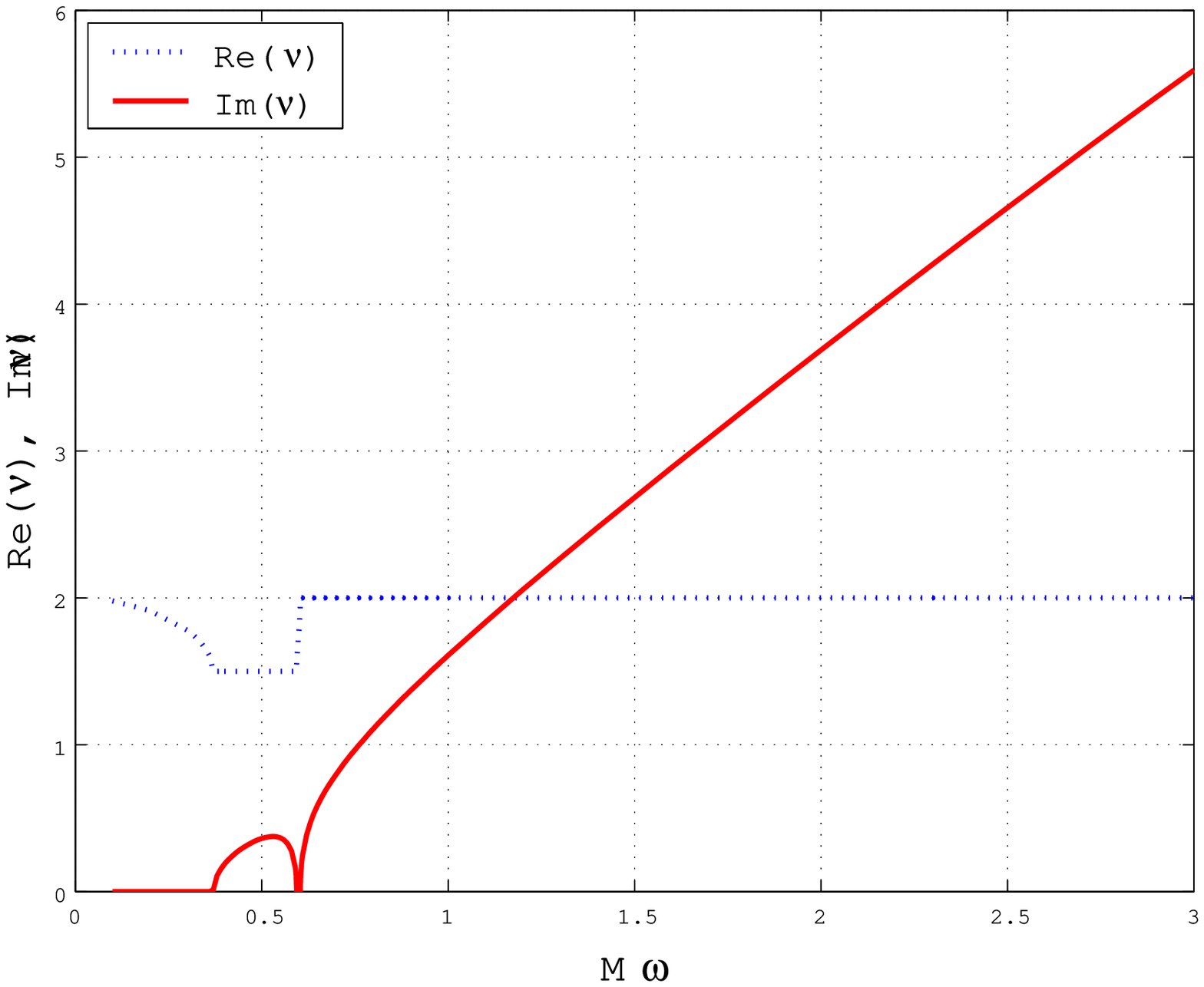}}
\caption{
The values of the real and imaginary parts of $\nu$ as functions
of $M\omega$ in the case $s=-2$, $l=m=2$ and $q=0$. 
}
\label{fig:nu-l2m2q0s2m}
}
\hfil
\parbox{\halftext}{
\centerline{\includegraphics[width=6.5cm,height=6cm]{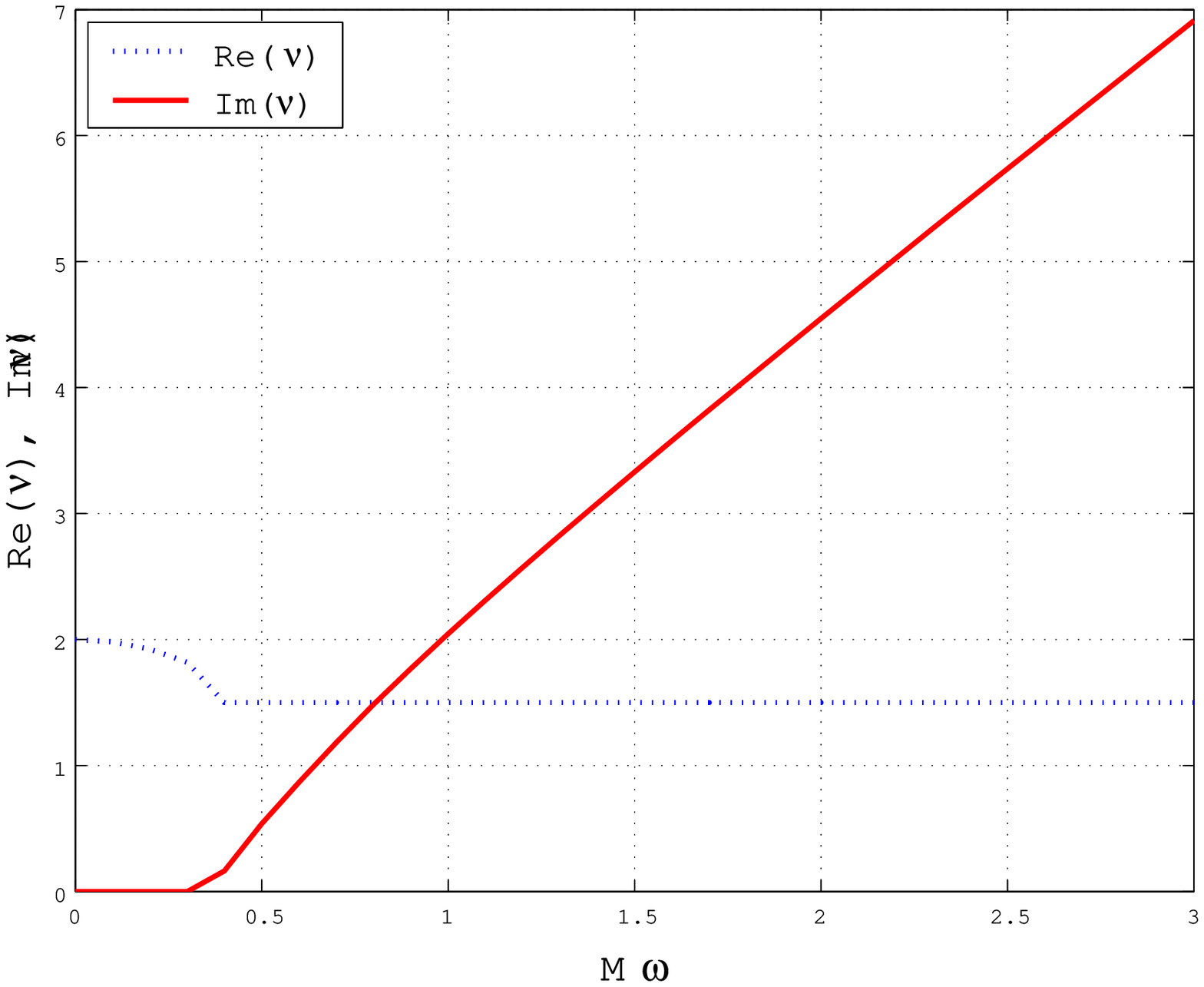}}
\caption{
The values of the real and imaginary parts of $\nu$ as functions
of $M\omega$ in the case $s=-2$, $l=m=2$ and $q=0.99$. 
}
\label{fig:nu-l2m2q099s2m}
}
\centerline{\includegraphics[width=6.5cm,height=6cm]{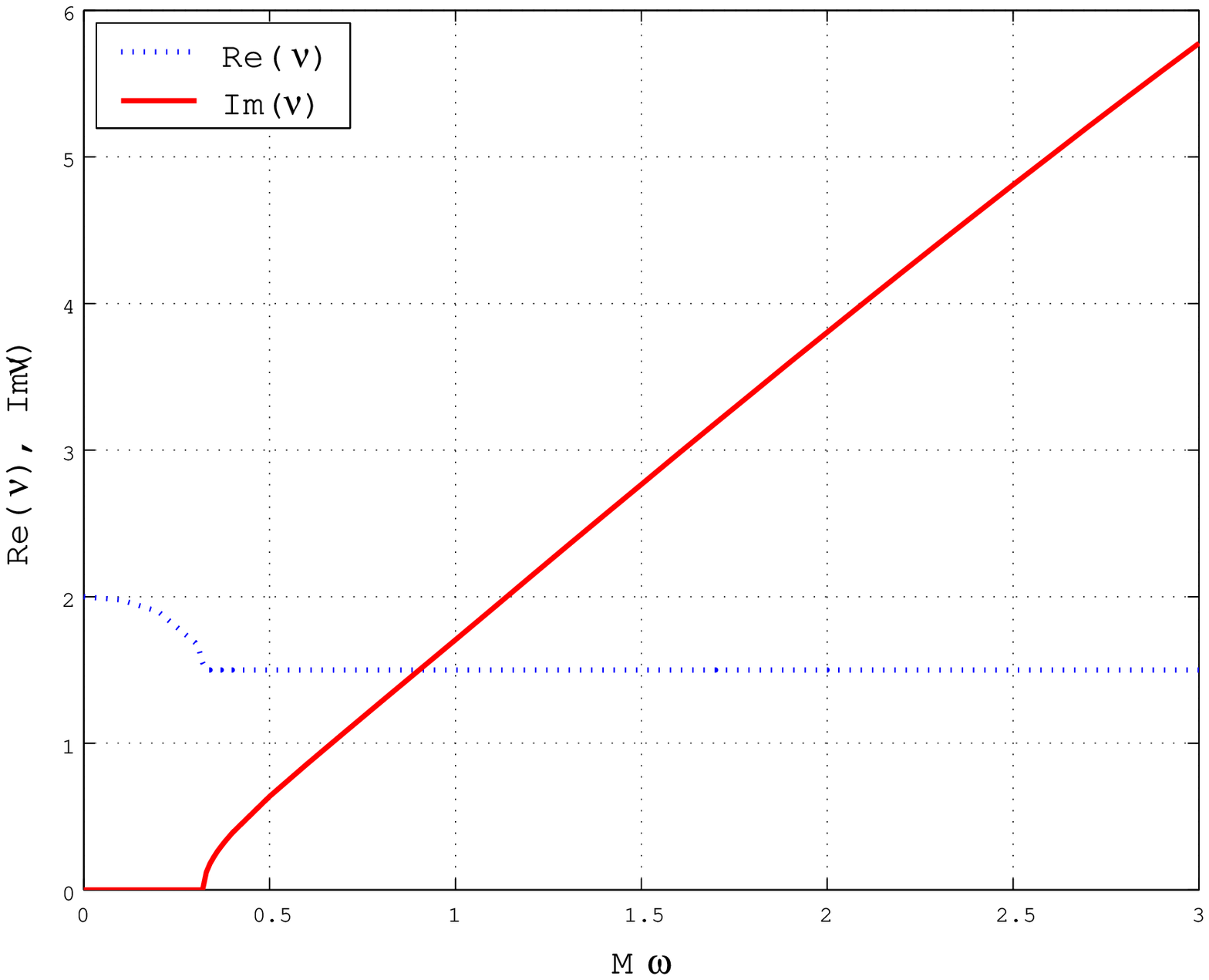}}
\caption{
The values of the real and imaginary parts of $\nu$ as functions
of $M\omega$ in the case $s=-2$, $l=m=2$ and $q=-0.99$. 
}
\label{fig:nu-l2m2q099ms2m}
\end{figure}

\begin{figure}[t]
\parbox{\halftext}{
\centerline{\includegraphics[width=6.5cm,height=6cm]{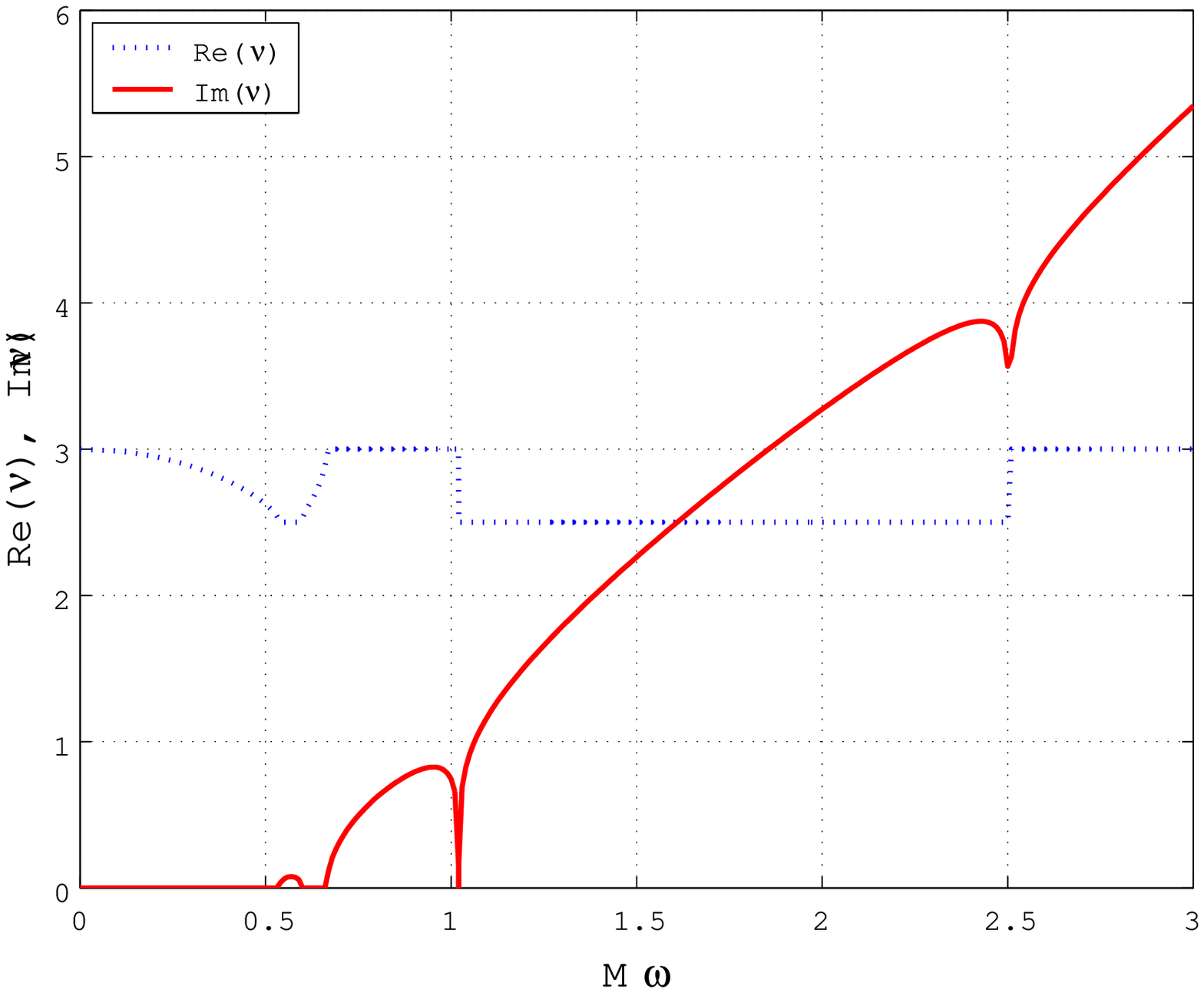}}
\caption{
The values of the real and imaginary parts of $\nu$ as functions
of $M\omega$ in the case $s=-2$, $l=m=3$ and $q=0$. 
}
\label{fig:nu-l3m3q0s2m}
}
\hfil
\parbox{\halftext}{
\centerline{\includegraphics[width=6.5cm,height=6cm]{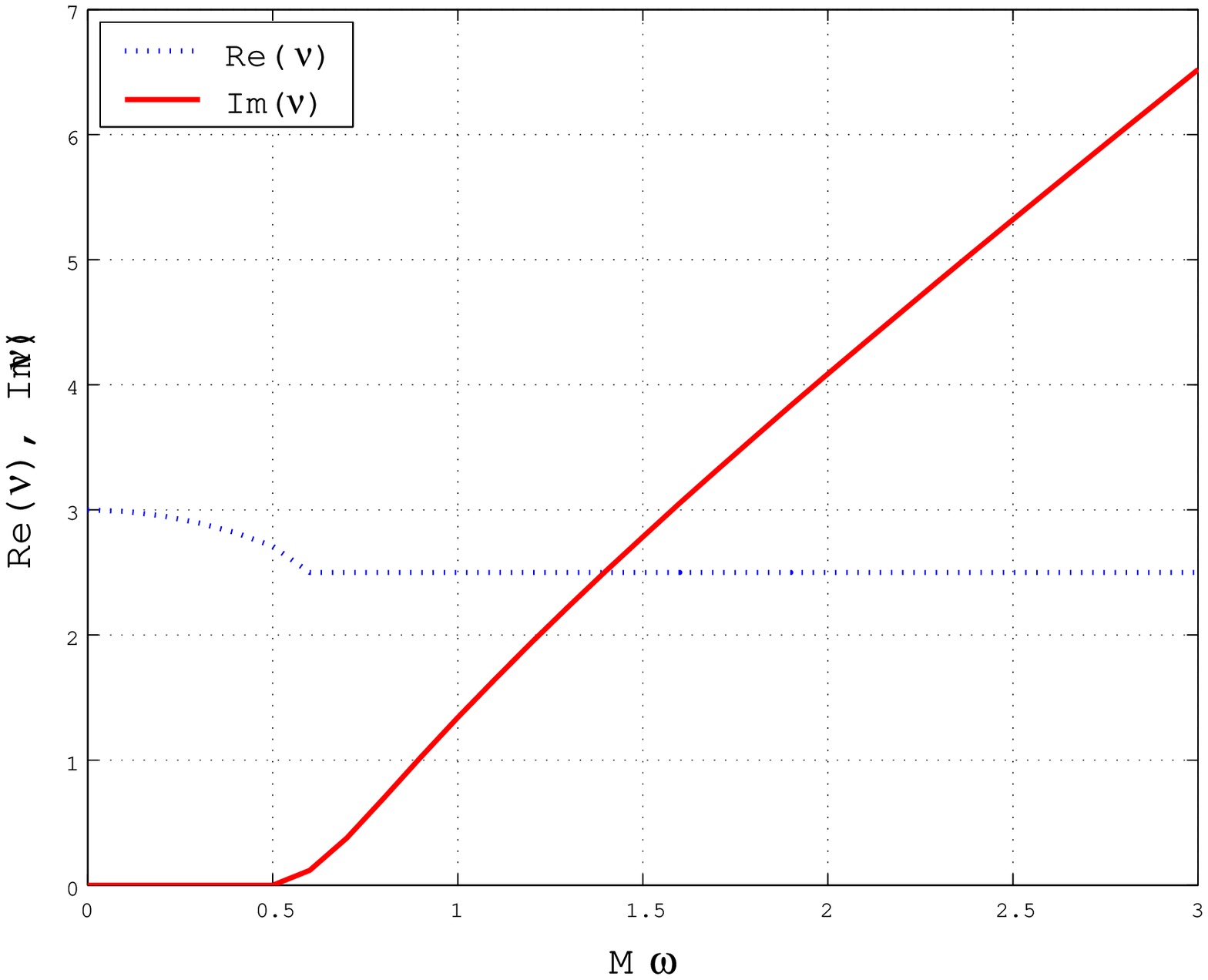}}
\caption{
The values of the real and imaginary parts of $\nu$ as functions
of $M\omega$ in the case $s=-2$, $l=m=3$ and $q=0.99$. 
}
\label{fig:nu-l3m3q099s2m}
}
\centerline{\includegraphics[width=6.5cm,height=6cm]{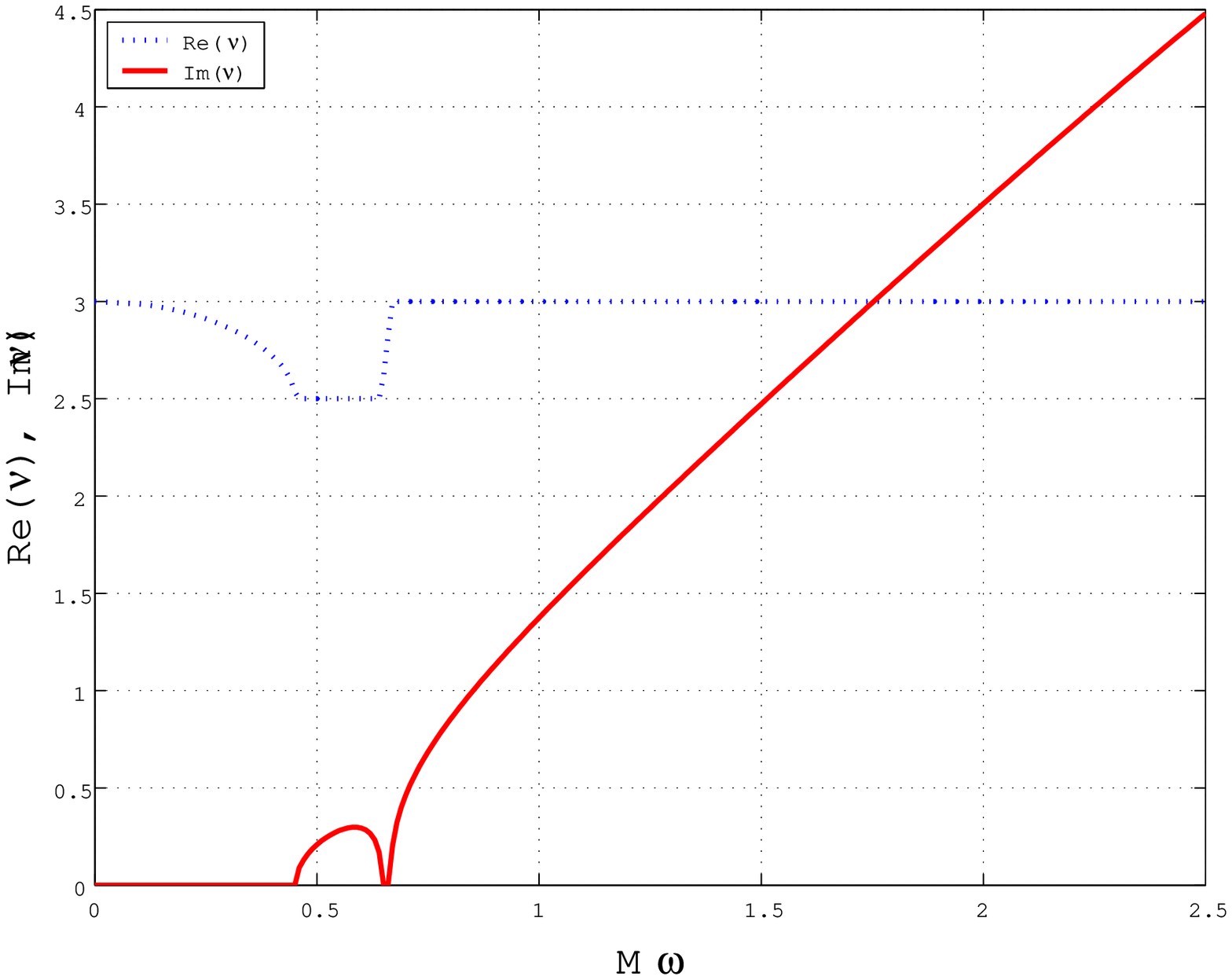}}
\caption{
The values of the real and imaginary parts of $\nu$ as functions
of $M\omega$ in the case $s=-2$, $l=m=3$ and $q=-0.99$. 
}
\label{fig:nu-l3m3q099ms2m}
\end{figure}

\begin{figure}[t]
\parbox{\halftext}{
\centerline{\includegraphics[width=6.5cm,height=6cm]{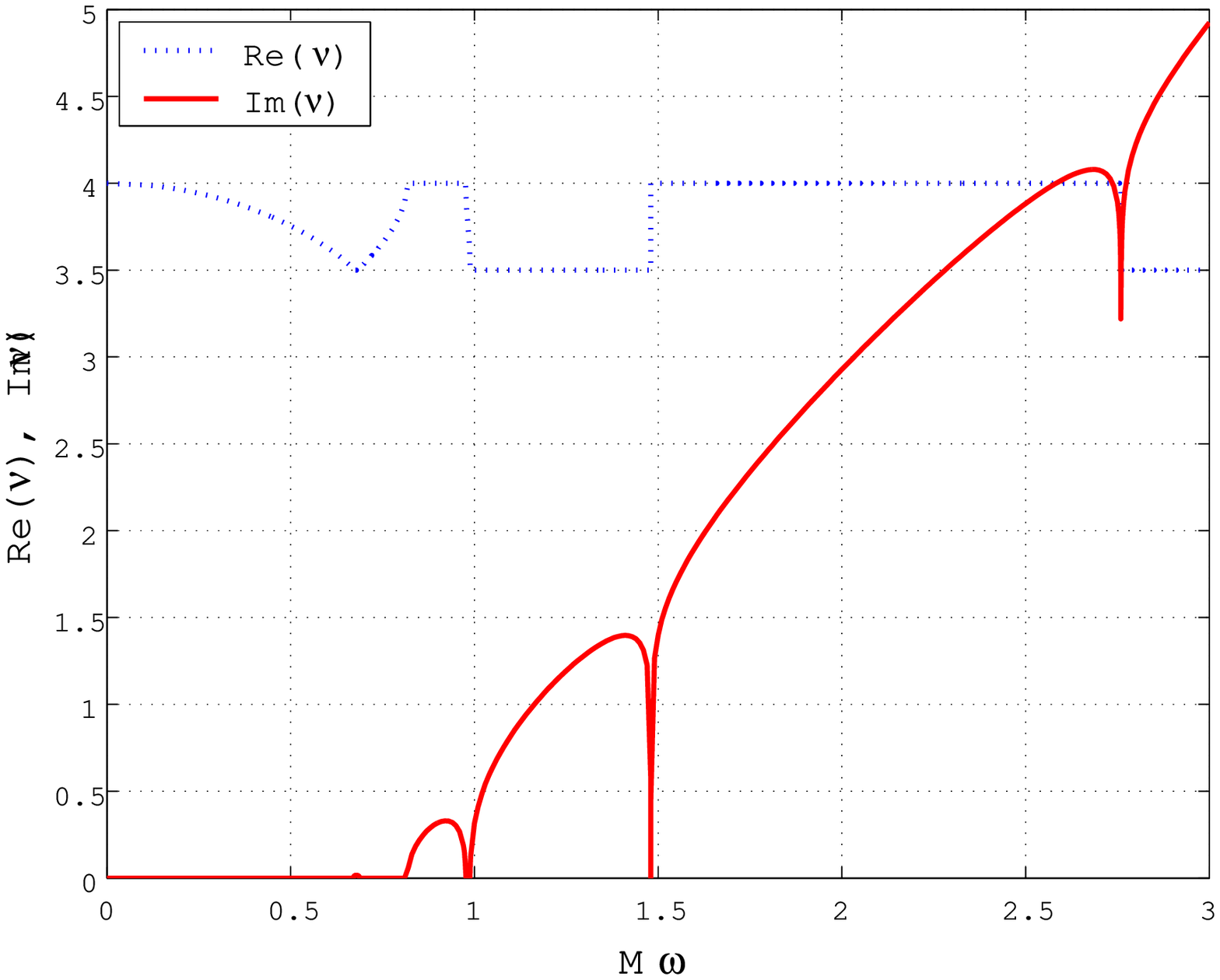}}
\caption{
The values of the real and imaginary parts of $\nu$ as functions
of $M\omega$ in the case $s=-2$, $l=m=4$ and $q=0$. 
}
\label{fig:nu-l4m4q0s2m}
}
\hfil
\parbox{\halftext}{
\centerline{\includegraphics[width=6.5cm,height=6cm]{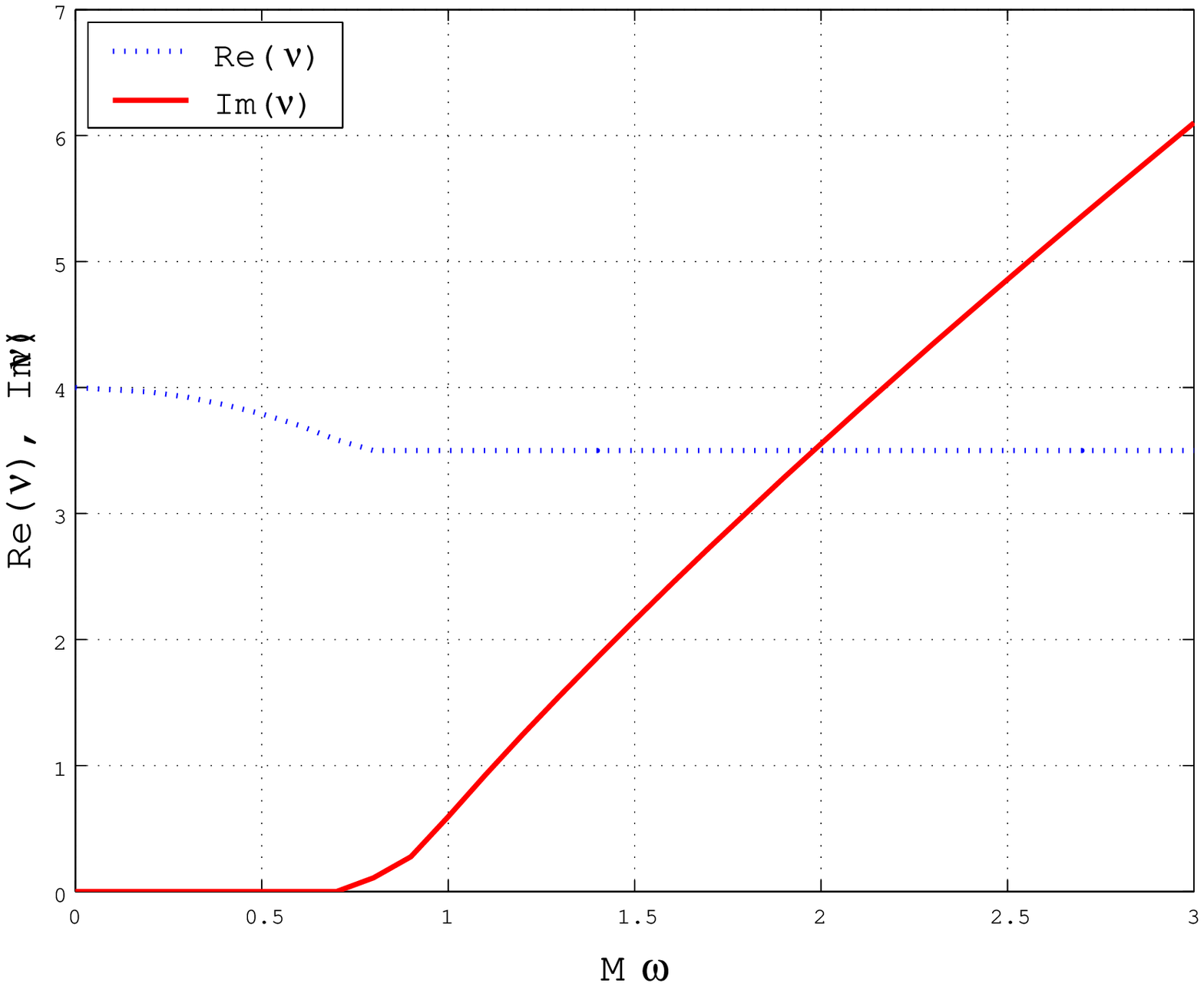}}
\caption{
The values of the real and imaginary parts of $\nu$ as functions
of $M\omega$ in the case $s=-2$, $l=m=4$ and $q=0.99$. 
}
\label{fig:nu-l4m4q099s2m}
}
\centerline{\includegraphics[width=6.5cm,height=6cm]{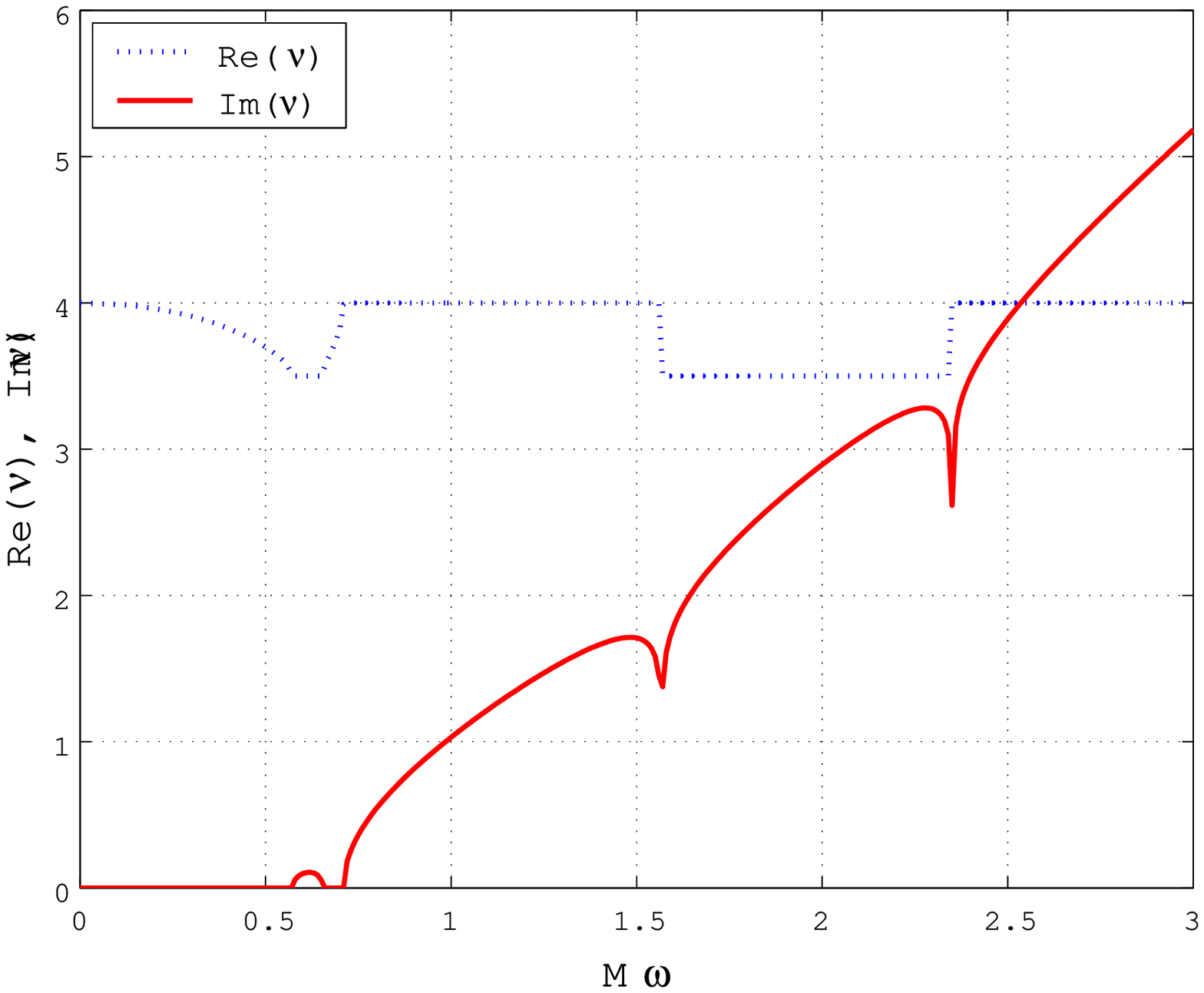}}
\caption{
The values of the real and imaginary parts of $\nu$ as functions
of $M\omega$ in the case $s=-2$, $l=m=4$ and $q=-0.99$. 
}
\label{fig:nu-l4m4q099ms2m}
\end{figure}

\begin{figure}[t]
\parbox{\halftext}{
\centerline{\includegraphics[width=6.5cm,height=6cm]{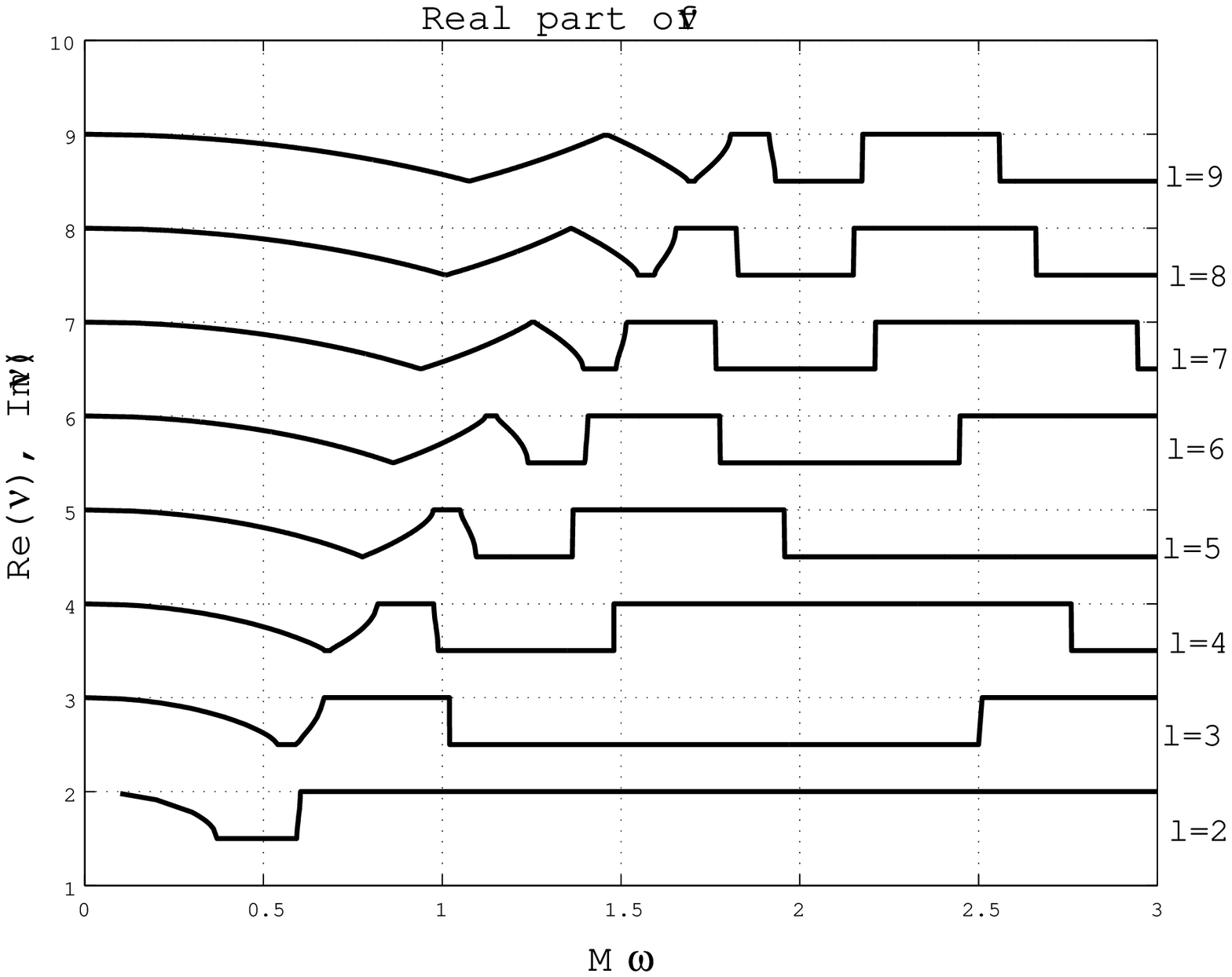}}
\caption{
The values of the real parts of $\nu$ as functions
of $M\omega$ in the cases $l=2,3,\cdots ,9$, $s=-2$ and $q=0$. 
}
\label{fig:nu-l2l9q0s2m_realpart}
}
\hfil
\parbox{\halftext}{
\centerline{\includegraphics[width=6.5cm,height=6cm]{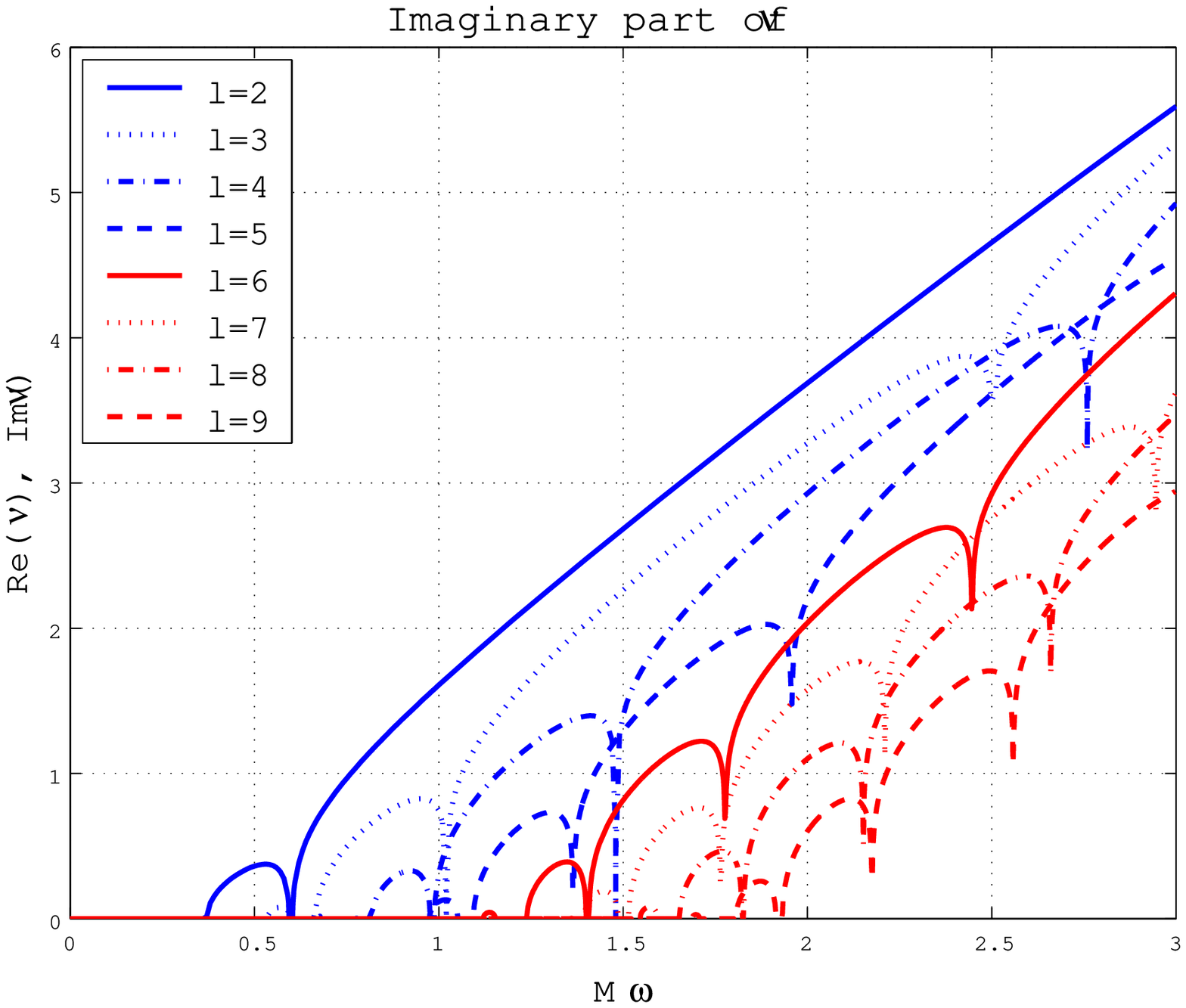}}
\caption{
The values of the imaginary parts of $\nu$ as functions
of $M\omega$ in the cases from $l=2,3,\cdots ,9$, $s=-2$ and $q=0$. 
}
\label{fig:nu-l2l9q0s2m_imagpart}
}
\end{figure}

In Figs. \ref{fig:nu-l2m2q0s2m}--\ref{fig:nu-l4m4q099ms2m},
we plot the real and imaginary parts of $\nu$ for $l=m=2,3,4$ and $q=0,\pm 0.99$. 
In Figs. \ref{fig:nu-l2l9q0s2m_realpart} and \ref{fig:nu-l2l9q0s2m_imagpart},
we plot the real and imaginary parts of $\nu$ for $l=2,\dots ,9$ in the case $q=0$. 
We find that the properties of $\nu$ are very similar to those in the case $l=2$, 
although there are several quantitative differences. 
For example, in the case $l=5$, in 
Figs. \ref{fig:nu-l2l9q0s2m_realpart} and \ref{fig:nu-l2l9q0s2m_imagpart},
${\rm Re}(\nu)$ reaches $4.5$ at $M\omega=0.776$, and a complex solution appears.
However, this complex solution exists only in a very small interval of $M\omega$,
$0.776\leq M\omega\leq 0.778$, and a real solution appears again for $M\omega>0.778$. 

In Tables \ref{tab:nu-s0l2m2q0}--\ref{tab:nu-s0l2m2q-099}, we list the values of 
$\nu$ for various $M\omega$ in the cases $s=0$, $l=m=2$ and $q=0,\pm 0.99$. 
We find behavior of $\nu$ very similar to that in the case $s=-2$. 

Here, we briefly explain our numerical method of searching for a solution of $g_0(z)=0$.
When we search for a real solution, it should be sufficient, in principle, to search 
the region $l-1/2 < x < l$, because of the symmetry property of the solution mentioned above.
However, in some cases, the function $g_0(x)$ becomes very steep around the solution
in that region. In such cases, it is often easier to search for the solution
in the region $l-1<x<l-1/2$. We thus search the region $l-1<x<l$. 
We discard solutions at integer and half-integer values, $x=l-1, l-1/2, l$.

When we do not find any solutions on the real $z$ axis, we search for a solution 
in the imaginary direction. We consider only half-integer and integer values of ${\rm Re}(z)$,
and search in the imaginary direction of the $z$-plane. 
Thus, the problem becomes a simple one-dimensional search for a root of the complex 
equation $g_0(z)=0$. 
Further, we utilize a very useful property of $g_0(z)$: The function $g_0(z)$ 
(with complex $z$) is real when $z=-1/2+iy$ (with real variable $y$). 
This is derived using the fact that $\beta_k=\beta_{-k}^*$ and 
$\alpha_k \gamma_{k+1}=(\alpha_{-k-1}\gamma_{-k})^*$ (where ${}^*$ denotes 
complex conjugation, and $k$ is an arbitrary integer). 
Thus, when we search for a solution with half-integer real part, 
it is convenient to search at ${\rm Re}(z)=-1/2$, since when $\nu$ is a solution, 
$\nu+k$ (with $k$ an arbitrary integer) is also a solution. 
For a solution with integer real part, we search for a solution which simultaneously satisfies
${\rm Re}(g_0(z))=0$ and ${\rm Im}(g_0(z))=0$.
In particular, we search for the real part ${\rm Re}(z)=1$ irrespective of $l$, because
the function $g_0(z)$ becomes steeper around the solution 
as ${\rm Re}(z)$ becomes larger, and it becomes more difficult to find the solution
in such cases. 

{\tiny
\begin{table}[t]
\caption{Table of $\nu$ for various values of $M\omega$ in the case $s=0$, $l=m=2$ and $q=0$. }
\begin{center}
\begin{tabular}{c|ll}
\hline \hline 
$M\omega$ & ${\rm Re}(\nu)$ & ${\rm Im}(\nu)$ \\ \hline 
0.1000 & 1.9848275014852 & 0.0000000000000\\
0.2000 & 1.9376748241581 & 0.0000000000000\\
0.3000 & 1.8522206563655 & 0.0000000000000\\
0.4000 & 1.7069234201112 & 0.0000000000000\\
0.4100 & 1.6861876869380 & 0.0000000000000\\
0.4200 & 1.6631943548601 & 0.0000000000000\\
0.4300 & 1.6369668328189 & 0.0000000000000\\
0.4400 & 1.6053426169451 & 0.0000000000000\\
0.4500 & 1.5608220358616 & 0.0000000000000\\
0.4600 & 1.5000000000000 & 0.0580486824445\\
0.4700 & 1.5000000000000 & 0.1001804231496\\
0.4800 & 1.5000000000000 & 0.1273338678120\\
0.4900 & 1.5000000000000 & 0.1476523972303\\
0.5000 & 1.5000000000000 & 0.1633287233824\\
0.5100 & 1.5000000000000 & 0.1751890776086\\
0.5200 & 1.5000000000000 & 0.1834763247507\\
0.5300 & 1.5000000000000 & 0.1880516714671\\
0.5400 & 1.5000000000000 & 0.1883986754070\\
0.5500 & 1.5000000000000 & 0.1834683228152\\
0.5600 & 1.5000000000000 & 0.1712231357384\\
0.5700 & 1.5000000000000 & 0.1472113005187\\
0.5800 & 1.5000000000000 & 0.0974051775175\\
0.5900 & 1.5967266037405 & 0.0000000000000\\
0.6000 & 1.6978839941760 & 0.0000000000000\\
0.6040 & 1.7370771353709 & 0.0000000000000\\
0.6080 & 1.7796485136225 & 0.0000000000000\\
0.6120 & 1.8295192010822 & 0.0000000000000\\
0.6160 & 1.8982359899500 & 0.0000000000000\\
0.6200 & 2.0000000000000 & 0.0881232882027\\
0.6600 & 2.0000000000000 & 0.4182708018765\\
0.7000 & 2.0000000000000 & 0.5798590491029\\
0.8000 & 2.0000000000000 & 0.8682632610900\\
0.9000 & 2.0000000000000 & 1.0958382861914\\
1.0000 & 2.0000000000000 & 1.2916407295318\\
1.1000 & 2.0000000000000 & 1.4596540676691\\
1.2000 & 2.0000000000000 & 1.5871735288388\\
1.3000 & 2.0000000000000 & 1.5757480036493\\
1.4000 & 1.5000000000000 & 1.8980859589471\\
1.6000 & 1.5000000000000 & 2.5246610689374\\
1.8000 & 1.5000000000000 & 3.0134306373653\\
2.0000 & 1.5000000000000 & 3.4652896348154\\
2.2000 & 1.5000000000000 & 3.8988668745912\\
2.4000 & 1.5000000000000 & 4.3212911554702\\
2.6000 & 1.5000000000000 & 4.7360933390574\\
2.8000 & 1.5000000000000 & 5.1452723630699\\
3.0000 & 1.5000000000000 & 5.5500484659006\\
\hline
\end{tabular}
\end{center}
\label{tab:nu-s0l2m2q0}
\end{table}
}

{\tiny
\begin{table}[t]
\caption{Table of $\nu$ for various values of $M\omega$ in the case $s=0$, $l=m=2$ and $q=0.99$. }
\begin{center}
\begin{tabular}{c|ll}
\hline \hline 
$M\omega$ & ${\rm Re}(\nu)$ & ${\rm Im}(\nu)$ \\ \hline 
0.10 & 1.9855006468982 & 0.0000000000000\\
0.20 & 1.9433423104280 & 0.0000000000000\\
0.30 & 1.8738919793067 & 0.0000000000000\\
0.40 & 1.7745498379081 & 0.0000000000000\\
0.50 & 1.6249603205669 & 0.0000000000000\\
0.51 & 1.6016482354182 & 0.0000000000000\\
0.52 & 1.5720774742602 & 0.0000000000000\\
0.53 & 1.5140777829215 & 0.0000000000000\\
0.54 & 1.5000000000000 & 0.0683318312500\\
0.55 & 1.5000000000000 & 0.0970789832967\\
0.56 & 1.5000000000000 & 0.1187106486632\\
0.57 & 1.5000000000000 & 0.1367752823053\\
0.58 & 1.5000000000000 & 0.1526950899429\\
0.59 & 1.5000000000000 & 0.1672593963838\\
0.60 & 1.5000000000000 & 0.1809982771760\\
0.70 & 1.5000000000000 & 0.3387878153683\\
0.80 & 1.5000000000000 & 0.5844007364833\\
0.90 & 1.5000000000000 & 0.8212346456388\\
1.00 & 1.5000000000000 & 0.9679578324407\\
1.10 & 2.0000000000000 & 1.0719514201554\\
1.20 & 2.0000000000000 & 1.5606928653937\\
1.30 & 2.0000000000000 & 1.8822793013188\\
1.40 & 2.0000000000000 & 2.1581224329993\\
1.50 & 2.0000000000000 & 2.4081352814494\\
1.60 & 2.0000000000000 & 2.6392239009791\\
1.70 & 2.0000000000000 & 2.8536938783809\\
1.80 & 2.0000000000000 & 3.0506896346940\\
1.90 & 2.0000000000000 & 3.2246989210551\\
2.00 & 2.0000000000000 & 3.3561709666536\\
2.10 & 2.0000000000000 & 3.2904400721719\\
2.20 & 1.5000000000000 & 3.7368758127391\\
2.30 & 1.5000000000000 & 4.0811236068986\\
2.40 & 1.5000000000000 & 4.3637677117354\\
2.50 & 1.5000000000000 & 4.6216867037460\\
2.60 & 1.5000000000000 & 4.8652774867248\\
2.70 & 1.5000000000000 & 5.0989848264810\\
2.80 & 1.5000000000000 & 5.3251065286991\\
2.90 & 1.5000000000000 & 5.5449488222337\\
3.00 & 1.5000000000000 & 5.7592716343452\\
\hline
\end{tabular}
\end{center}
\label{tab:nu-s0l2m2q099}
\end{table}
}

{\tiny
\begin{table}[t]
\caption{Table of $\nu$ for various values of $M\omega$ in the case $s=0$, $l=m=2$ and $q=-0.99$. }
\begin{center}
\begin{tabular}{c|ll}
\hline \hline 
$M\omega$ & ${\rm Re}(\nu)$ & ${\rm Im}(\nu)$ \\ \hline 
0.10 & 1.9840980928918 & 0.0000000000000\\
0.20 & 1.9308382464262 & 0.0000000000000\\
0.30 & 1.8203733727014 & 0.0000000000000\\
0.31 & 1.8041051082066 & 0.0000000000000\\
0.32 & 1.7863405245509 & 0.0000000000000\\
0.33 & 1.7667869072726 & 0.0000000000000\\
0.34 & 1.7450236585402 & 0.0000000000000\\
0.35 & 1.7204062673298 & 0.0000000000000\\
0.36 & 1.6918506722768 & 0.0000000000000\\
0.37 & 1.6572410906180 & 0.0000000000000\\
0.38 & 1.6110789879720 & 0.0000000000000\\
0.39 & 1.5000000000000 & 0.0186032751635\\
0.40 & 1.5000000000000 & 0.1154238436610\\
0.50 & 1.5000000000000 & 0.3965366549457\\
0.60 & 1.5000000000000 & 0.5628860942422\\
0.70 & 1.5000000000000 & 0.6786359420169\\
0.80 & 1.5000000000000 & 0.6765122702806\\
0.90 & 2.0000000000000 & 0.9125809354487\\
1.00 & 2.0000000000000 & 1.2541180907051\\
1.10 & 2.0000000000000 & 1.5100321536697\\
1.20 & 2.0000000000000 & 1.7355022863878\\
1.30 & 2.0000000000000 & 1.9421454082317\\
1.40 & 2.0000000000000 & 2.1324040070446\\
1.50 & 2.0000000000000 & 2.3031629738421\\
1.60 & 2.0000000000000 & 2.4408891644519\\
1.70 & 2.0000000000000 & 2.4732101098580\\
1.80 & 1.5000000000000 & 2.7185951268120\\
1.90 & 1.5000000000000 & 3.0969911535975\\
2.00 & 1.5000000000000 & 3.3873306639375\\
2.10 & 1.5000000000000 & 3.6492406015533\\
2.20 & 1.5000000000000 & 3.8963211129597\\
2.30 & 1.5000000000000 & 4.1339652371712\\
2.40 & 1.5000000000000 & 4.3648798107882\\
2.50 & 1.5000000000000 & 4.5906202696084\\
2.60 & 1.5000000000000 & 4.8121617032199\\
2.70 & 1.5000000000000 & 5.0301522240344\\
2.80 & 1.5000000000000 & 5.2450395718077\\
2.90 & 1.5000000000000 & 5.4571399180362\\
3.00 & 1.5000000000000 & 5.6666774973656\\
\hline
\end{tabular}
\end{center}
\label{tab:nu-s0l2m2q-099}
\end{table}
}

\section{Gravitational wave luminosity}
\label{sec:luminosity}

Since the role of $\nu$ is to match two minimal solutions in the limits $n\rightarrow\pm\infty$ 
of the recurrence relation, it is evident that 
once we find a solution of Eq. (\ref{eq:determine_nu}),
it can be used to generate homogeneous solutions irrespective of whether 
it is real or complex. 
In this section, we confirm this by using 
the luminosity of gravitational waves radiated by a 
particle in a circular, equatorial orbit around a Kerr black hole. 
The complete formulas are given in Appendix A of paper I. 
We compare the numerical data with those computed with the numerical 
integration method of Kennefick. \footnote{The numerical data were kindly 
provided to us by D. Kennefick and are based on the work presented in Ref. \citen{GK}.} 

The innermost stable circular orbit (ISCO) around a Kerr black hole is 
given by 
\begin{eqnarray}
\label{eq:risco}
r_{\rm ISCO}&=&M[3+Z_2\mp\sqrt{(3-Z_1)(3+Z_1+2Z_2)}],\\
\label{eq:riscoZ1}
Z_1&=&1+(1-q^2)^{1/3}[(1+q)^{1/3}+(1-q)^{1/3}],\\
\label{eq:riscoZ2}
Z_2&=&\sqrt{3q^2+Z_1^2},
\end{eqnarray}
where the upper and lower signs refer to the cases $q>0$ and
$q<0$, respectively. 
The orbital angular frequency of a circular orbit is given by 
\begin{eqnarray}
\Omega=\frac{M^{1/2}}{r^{3/2}(1+q(M/r)^{3/2})}.
\end{eqnarray}
The angular frequency of gravitational waves corresponding to 
the harmonic $m$ is given by $\omega=m\Omega$. 
For $q>0$, the orbital radius of the ISCO becomes smaller than that in the 
Schwarzschild case, and it approaches the event horizon as $q\rightarrow 1$. 
Thus, $M\omega$ around the ISCO can be larger than $(M\omega)_{\rm max}$. 

In Table \ref{tab:flux-compare_kenn}, we list the numerical data for the gravitational wave luminosity
induced by a particle in a circular orbit in the case that $M\omega$ is larger than $(M\omega)_{\rm max}$. 
We find that the numerical data computed using a complex $\nu$ agree with those computed 
using the numerical integration method to about 5 or 6 digits. 
This is approximately the same degree of accuracy of the 
numerical integration. 
This demonstrates the validity of using a complex $\nu$ to produce the 
homogeneous solutions of the Teukolsky equation. 

\begin{table}
\caption{Comparison of the energy fluxes obtained using 
the numerical integration method of Kennefick
and the method presented here in the case $r_0=1.55M$ and $q=0.99$, with complex $\nu$.}
\begin{center}
\begin{tabular}{|c|c|c|c|c|c|c|}
\hline \hline 
$l$&$\mid m\mid$&Re($\nu$)&Im($\nu$)&{\rm Numerical integration}&{\rm This paper}&{\rm Relative error} \\ \hline \hline 
$ 2$&$ 2$&$1.5$&$1.1374192131794$ &$3.568050135\times 10^{-2}$&$3.568033154338851\times 10^{-2}$&$4.76\times 10^{-6}$\\
$ 3$&$ 3$&$2.5$&$1.4249576682707$ &$2.152962111\times 10^{-2}$&$2.152959342790158\times 10^{-2}$&$1.29\times 10^{-6}$\\
$ 4$&$ 4$&$3.5$&$1.7677955367662$ &$1.230541176\times 10^{-2}$&$1.230541952573211\times 10^{-2}$&$6.31\times 10^{-7}$\\
$ 5$&$ 4$&$4.5$&$0.3735768429955$ &$1.933923538\times 10^{-5}$&$1.933924400940079\times 10^{-5}$&$4.46\times 10^{-7}$\\
$ 5$&$ 5$&$4.5$&$2.1437000387424$ &$7.259841388\times 10^{-3}$&$7.259849874157195\times 10^{-3}$&$1.17\times 10^{-6}$\\
$ 6$&$ 5$&$5.5$&$0.8032470628900$ &$1.536518877\times 10^{-5}$&$1.536520289551414\times 10^{-5}$&$9.19\times 10^{-7}$\\
$ 6$&$ 6$&$5.5$&$2.5395166388946$ &$4.404590776\times 10^{-3}$&$4.404599359654937\times 10^{-3}$&$1.95\times 10^{-6}$\\
$ 7$&$ 6$&$6.5$&$1.2521347909128$ &$1.148793520\times 10^{-5}$&$1.148795883734041\times 10^{-5}$&$2.06\times 10^{-6}$\\
$ 7$&$ 7$&$6.5$&$2.9481395894691$ &$2.726943363\times 10^{-3}$&$2.726949666682515\times 10^{-3}$&$2.31\times 10^{-6}$\\
$ 8$&$ 7$&$7.5$&$1.6997166387020$ &$8.267139284\times 10^{-6}$&$8.267168128593364\times 10^{-6}$&$3.49\times 10^{-6}$\\
$ 8$&$ 8$&$7.5$&$3.3654788843469$ &$1.713109264\times 10^{-3}$&$1.713110726903289\times 10^{-3}$&$8.54\times 10^{-7}$\\
$ 9$&$ 7$&$8.5$&$0.3781387034254$ &$1.650193407\times 10^{-7}$&$1.650193971262395\times 10^{-7}$&$3.42\times 10^{-7}$\\
$ 9$&$ 8$&$8.5$&$2.1467598463840$ &$5.800131555\times 10^{-6}$&$5.800152000099428\times 10^{-6}$&$3.53\times 10^{-6}$\\
$ 9$&$ 9$&$8.5$&$3.7890216134824$ &$1.087833559\times 10^{-3}$&$1.087830324625970\times 10^{-3}$&$2.97\times 10^{-6}$\\
$10$&$ 8$&$9.5$&$0.8445521091574$ &$1.423932484\times 10^{-7}$&$1.423939557356283\times 10^{-7}$&$4.97\times 10^{-6}$\\
$10$&$ 9$&$9.5$&$2.5936449978532$ &$3.998328655\times 10^{-6}$&$3.998328388790399\times 10^{-6}$&$6.66\times 10^{-8}$\\
$10$&$10$&$9.5$&$4.2171390613851$ &$6.963970144\times 10^{-4}$&$6.963929698290593\times 10^{-4}$&$5.81\times 10^{-6}$\\
\hline \hline
\end{tabular}
\end{center}
\label{tab:flux-compare_kenn}
\end{table}

\begin{figure}[t]
\centerline{\includegraphics[width=6.5cm,height=6cm]{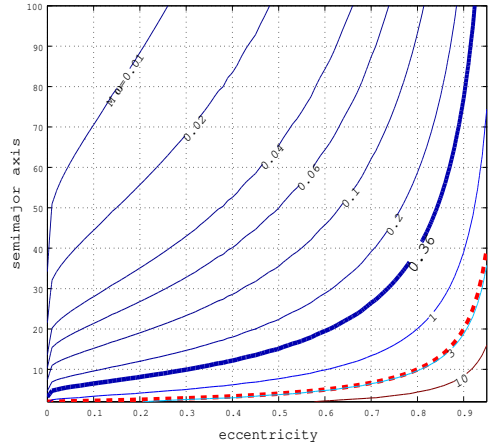}}
\caption{
The contours of the values of $M\omega(=Mn\omega_0)$ that must be included 
in the evaluation of the total power in order to obtain the the relative error less than $10^{-5}$. 
Here, $M=m_1+m_2$. 
The thick curve corresponds to the contour for $M\omega=0.36$, which 
is equal to $(M\omega)_{\rm max}$, in the case $l=m=2$ and $q=0$, 
above which $\nu$ becomes complex. 
The region under thick dashed curve should be excluded from consideration, 
because there, the pericenter distance is less than $2M$. 
Thus, in the region between the thick curve and the thick dashed curve, 
the parameter $\nu$ can have a complex value. 
}
\label{fig:neededomega}
\end{figure}

In the case of eccentric orbits, there are infinite number of higher harmonics
for each $l,m$. Thus, in principle, there are many harmonics 
for which $M\omega$ is larger than $(M\omega)_{\rm max}$. 
Here, we estimate the range of harmonics needed to 
express the total power of the gravitational waves with the relative error 
less than $10^{-5}$ using the Newtonian, 
quadrupole approximation, in the case of a non-spinning binary. 
The power radiated in $n$th harmonics is given by \cite{PetersMathews}
\begin{eqnarray}
P(n)=\frac{32}{5}\frac{m_1^2m_2^2(m_1+m_2)}{r_a^5}g(n,e),
\end{eqnarray}
where $m_1$ and $m_2$ are the masses of the two stars forming the binary, $e$ is the eccentricity, 
$r_a$ is the semi-major axis of the orbit, and we have
\begin{eqnarray}
g(n,e)&=&\frac{n^4}{32}\Bigg\{\left[J_{n-2}(ne)-2eJ_{n-1}(ne)+\frac{2}{n}J_n(ne)
+2eJ_{n+1}(ne)-J_{n+2}(ne)\right]^2\nonumber\\
&&+(1-e^2)[J_{n-2}(ne)-2J_n(ne)+J_{n+2}(ne)]^2+\frac{4}{3n^2}(J_n(ne))^2\Bigg\},
\end{eqnarray}
where $J_{n}(z)$ is the Bessel function. 
The angular frequency for each harmonic is given by $\omega=n\omega_0$, with
$\omega_0=(m_1+m_2)^{1/2}/r_a^{3/2}$.
In Fig. \ref{fig:neededomega}, we plot the contours of 
the values of $M\omega$ that must
be included in the total power as functions of $e$ and $r_a$. 
We find that for small $r_a$, there is a relatively wide range of values of $e$ for which 
the required value of $M\omega$ exceeds $(M\omega)_{\rm max}=0.36$ 
in the case $l=m=2$ and $q=0$. 
In such cases, complex $\nu$ must be employed in order to evaluate 
the gravitational radiation. 
Of course, we should keep in mind that, 
because this is based on the Newtonian, quadrupole approximation, 
this result will be changed quantitatively in the case of fully relativistic treatment.

\section{Summary and discussion}
\label{sec:summary}

We have investigated the numerical properties of the solutions of 
the continued fraction equation derived by Leaver and Mano et al. 
in the formalism of the 
analytic representation of the homogeneous solution of the Teukolsky 
equation. 
We have found that, for each $s,l,m$ and $q$, the solution $\nu$ takes 
a real or complex value, depending on the value of $\omega$. 
We have also found that, when $\nu$ is complex, 
its real part is always an integer or half integer. 
Although there are certain regularities in the behavior of the solution, 
we have not determined how such behavior is controlled by the continued fraction equation.
We hope to investigate the behavior of $\nu$ analytically
in the future. 

In order to confirm the existence of complex $\nu$, 
we have computed the gravitational wave luminosity emitted by a particle
in circular, equatorial orbit around a Kerr black hole in the case that 
$\nu$ is complex, and compared the results with those computed using 
a numerical integration method. We found that the two results are consistent. 
This shows clearly the validity of the existence of complex $\nu$. 
Even if $M\omega$ becomes larger than a certain critical value and a 
real $\nu$ does not exist, we can use the MST formalism by treating $\nu$ as 
a complex value. 
Thus, the region of $\omega$ in which the MST formalism can be used becomes 
much wider than in the case that we consider only real $\nu$, as in paper I.

There is, however, a limitation on the values of $\omega$ 
for which we can use the MST formalism for actual numerical computations. 
When $M\omega$ becomes very large (say $M\omega>5$ in the case $s=-2$, $l=2$ and $q=0$),
it becomes very difficult to evaluate the continued fraction 
equation (\ref{eq:determine_nu}) accurately because of the loss of accuracy. 
As $\omega$ becomes large, $\beta_n$ approaches 
$-(\alpha_n^\nu R_{n+1}+\gamma_n^\nu L_{n-1})$ 
irrespective of $\nu$. 
Thus, it becomes difficult to evaluate Eq. (\ref{eq:determine_nu}) numerically 
with high accuracy. We must solve this problem because $M\omega$ becomes 
very large in the cases of generic orbits, which are important for the data 
analysis for LISA. A possible solution is to derive the high frequency expansion of $\nu$.
If we cannot solve the problem, we may need to use numerical integration methods to evaluate the homogeneous 
solutions of the Teukolsky equation when $M\omega$ becomes very large.
However, we believe that the MST method supplemented by some numerical integration 
is a very powerful method for the numerical computation of the Teukolsky equation. 

Based on the analysis presented in this paper, 
we plan to apply the MST formalism to the evaluation of the gravitational 
wave luminosity in the cases of eccentric orbits and more generic, non-equatorial 
orbits, and to investigate the data analysis issue for LISA in the near future.

\section*{Acknowledgements}
We thank D. Kennefick, who kindly provided us with his numerical data 
for the gravitational wave luminosity computed with the code 
developed in his work \cite{GK}. 
We also thank M. Sasaki and F. Takahara for continuous encouragement. 
H. T. thanks K. S. Thorne for continuous encouragement and his hospitality
while he was at the California Institute of Technology, where a part of
this work was done, as a Zaigai Kenkyuin of Monbukagaku-sho. 
This work was supported in part by Monbukagaku-sho Grants-in-Aid
for Scientific Research (Nos. 14047214, 12640269 and 16540251).

\end{document}